\documentclass[%
 aip,
 amsmath,amssymb,
 reprint,%
]{revtex4-1}

\usepackage{graphicx}
\usepackage{dcolumn}
\usepackage{bm}

\usepackage[utf8]{inputenc}
\usepackage[T1]{fontenc}
\usepackage{mathptmx}
\usepackage{etoolbox}
\usepackage{physics}
\usepackage{caption}
\usepackage{subfloat}
\usepackage{subcaption}

\newcommand{\deu}[1]{\textnormal{D}\textsubscript{2}}
\newcommand{\neop}[1]{\textnormal{C}\textsubscript{5}\textnormal{H}\textsubscript{12}}

\makeatletter
\def\@email#1#2{%
 \endgroup
 \patchcmd{\titleblock@produce}
  {\frontmatter@RRAPformat}
  {\frontmatter@RRAPformat{\produce@RRAP{*#1\href{mailto:#2}{#2}}}\frontmatter@RRAPformat}
  {}{}
}%
\makeatother
\begin{document}

\preprint{AIP/123-QED}

\title[Nonlocal effects on Thermal Transport in MagLIF-Relevant Gaspipes on NIF]{Nonlocal effects on Thermal Transport in MagLIF-Relevant Gaspipes on NIF}
\author{R. Y. Lau}
 \email{ryan.lau@colorado.edu.}
\affiliation{University of Colorado Boulder, Boulder, USA}%
\author{D. J. Strozzi}%
\affiliation{Lawrence Livermore National Laboratory, P.O. Box 808, Livermore California 94551-0808, USA}
\author{M. Sherlock}%
\affiliation{Lawrence Livermore National Laboratory, P.O. Box 808, Livermore California 94551-0808, USA}
\author{M. Weis}%
\affiliation{Sandia National Laboratories, 1515 Eubank SE, Albuquerque, New Mexico 87123, USA}
\author{A. S. Joglekar}%
\affiliation{Ergodic LLC, 7511 Greenwood Avenue North, \#312 Seattle, WA 98103}
\author{W. A. Farmer}%

\affiliation{Lawrence Livermore National Laboratory, P.O. Box 808, Livermore California 94551-0808, USA}
\author{Y. Shi}
\affiliation{University of Colorado Boulder, Boulder, USA}
\author{J. R. Cary}
\affiliation{University of Colorado Boulder, Boulder, USA}

\date{\today}

\begin{abstract}
We present simulations of heat flow relevant to gaspipe experiments on the National Ignition Facility (NIF) to investigate kinetic effects on transport phenomena. \deu{} and neopentane (\neop{}) filled targets are used to study the laser preheat stage of a MagLIF scheme where an axial magnetic field is sometimes applied to the target. Simulations were done with the radiation-MHD code HYDRA with a collision-dominated fluid model and the Schurtz nonlocal electron thermal conduction model. Using the Schurtz model to evolve the electron temperature increased the heat front propagation of neopentane gas targets compared to a local model by limiting radial heat flow. This increases electron temperature near the axis, which decreases laser absorption. We find the effect of heat flow models on temperature profiles and laser propagation is modest. Beyond the Schurtz model, we utilize HYDRA to initialize plasma conditions for the Vlasov-Fokker-Planck K2 code.  We run K2 until a quasi-steady state is reached and examine the impact of kinetic effects on heat transport. Although axial heat flow is well predicted by fluid models, the fluid model consistently over predicts radial heat flow up to 150\% in regions with the largest temperature gradient of \deu{} filled gaspipes. On the other hand, the Schurtz nonlocal electron conduction model is found to be adequate for capturing kinetic heat flow in gaspipes. 
\end{abstract}

\maketitle

\section{\label{sec: intro} Introduction}
Recent experiments at the National Ignition Facility\cite{NIFref} (NIF) have been designed to study the preheat stage of a Magnetized Liner Inertial Fusion (MagLIF) scheme\cite{Slutz2012} which is a magneto-inertial confinement fusion concept developed by Sandia National Laboratories. As opposed to usual inertial confinement fusion (ICF) schemes of laser direct drive\cite{directdrive} or laser indirect drive\cite{indirectdrive}, which spherically compress and heat fusion fuel (usually deuterium or a mixture of deuterium and tritium) through the use of lasers, MagLIF utilizes a cylindrical geometry and compresses the liner through a pulsed power driver.
MagLIF has three stages involving the initial magnetization axially down the liner, a laser heating stage to condition the adiabat of the plasma, and a final compression stage of the plasma to reach self-heating conditions. The external, axial magnetic field reduces electron thermal loss from the hotspot, and increases alpha-particle heating. The second stage heats the fuel to several hundred eV. Without this preheat stage, fusion conditions cannot be reached by the magnetic field and compression alone.\cite{Slutz2012,Slutz2018,Sefkow2014,Harvey-Thompson2019,Gomez2014} Finally, the liner and fuel are compressed by radial $J\times B$ forces from axial current driven by the pulsed power system\cite{Cuneo2012}.

However, MagLIF at a fusion scale requires 20-30 kJ of laser energy to be coupled to the fuel which is not yet attainable at the Z facility\cite{Gomez2014, Gomez2020}. On the other hand, the NIF can utilize a single quad to deliver 30 kJ of 351 nanometer wavelength light to a cylindrical fuel cavity representative of a MagLIF target (gaspipe). Radial compression of the liner has not been done on the NIF however. MagLIF preheat experiments have been used to study laser energy coupling and liner mixing \cite{MagLIFexp}. We utilize the latest modeling of these experiments as a platform to study kinetic effects in heat flow\cite{Weis2025}. 
There has been a long history of radiation-hydrodynamic (rad-hydro) modeling on ICF experiments which utilize artificial flux limiters\cite{Shvarts1981,fluxLimiter,Goncharov2006,Goncharov2005,Boehly2011} to imitate kinetic heat flow effects to match experimental data. A common approach is to take the minimum between a local fluid model of heat flow (such as Spitzer-Härm\cite{Spitzer}, or Braginskii\cite{Braginskii} models), $Q_{SH}$, and a constant $f_L$ multiplied by the free-streaming limit $Q_{FS}=n_e T_e v_{th}$ for $v_{th} \equiv \sqrt{T_e/m_e}$.\cite{fluxLimiter,Davies2015} While this produces the desired effect of limiting the heat flow, it provides an unsatisfactory explanation of the physics at hand and fails to reproduce other kinetic effects. \cite{fluxLimiter,Boehly2011}. 
Although utilizing a more accurate model of heat flow could lead to a better understanding of modeling HED experiments in general, the computational costs to implement fully kinetic models on hydrodynamic timescales are extremely expensive. 
Therefore, there is great motivation to study the validity of available models depending on the plasma parameters of the problem. In this paper, we investigate when it is appropriate to use a local heat transport model, versus using a reduced nonlocal heat transport model, versus using a kinetic heat transport model.
\begin{figure} [h]
\centering 
\includegraphics[scale=0.5]{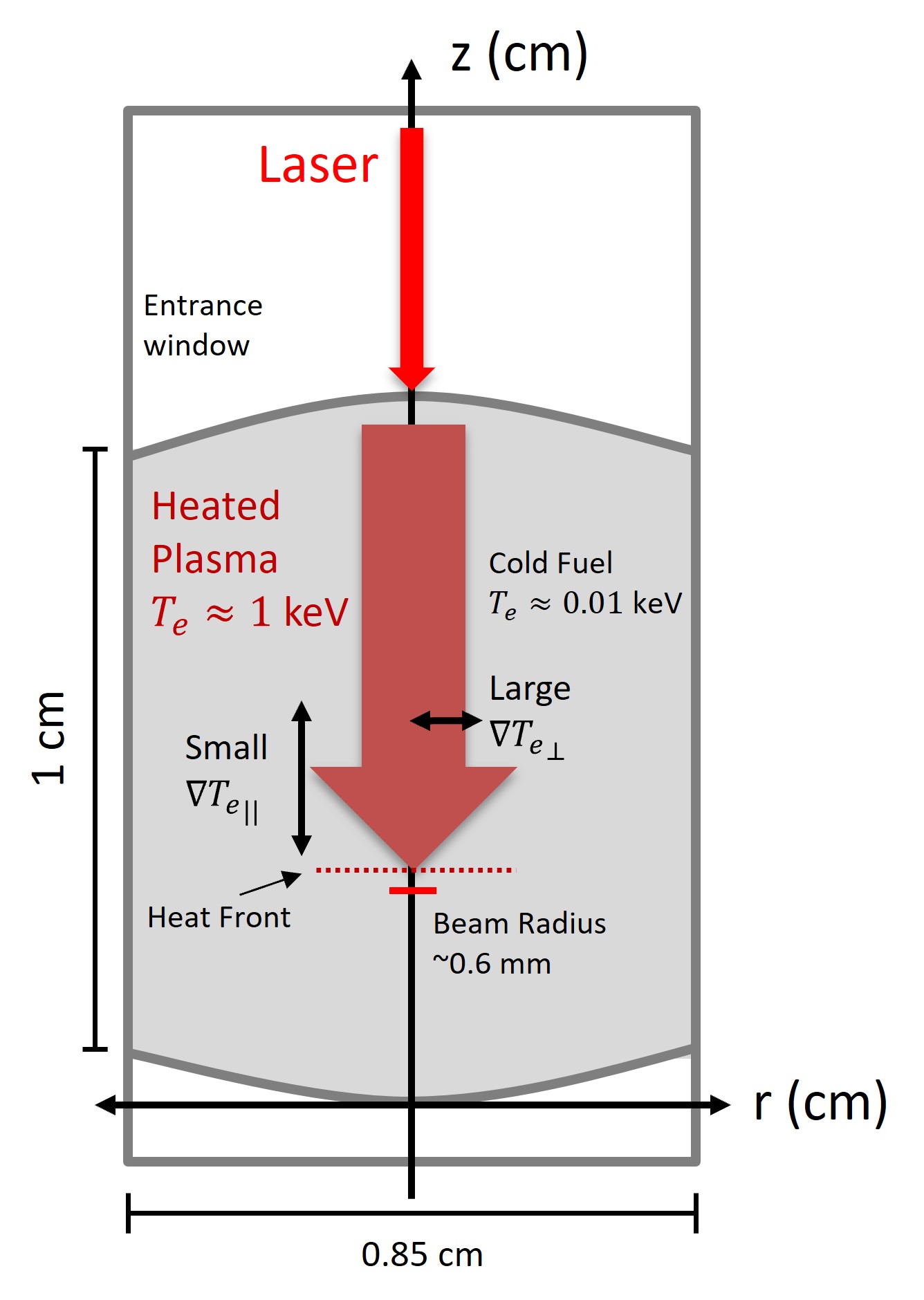} 
\caption{A simplified (not to scale) 2D $rz$ cross-section of the geometry of a gaspipe. Because the laser attenuates over a distance much longer in $z$ compared to $r$, the larger temperature gradients and the more nonlocal conditions occur in the radial direction. }
\label{fig: simpleOutline} 
\end{figure}

Nonlocality refers to transport phenomena that cannot be captured by a first-order expansion in the distribution function by a small parameter such as the Knudsen number Kn. An indication for when local fluid models of electron heat flow begin to fail\cite{Bell1981} is when Kn exceeds 1\% defined as
\begin{gather}
    \textrm{Kn} \equiv \frac{\lambda_{\textrm{col}}}{L_T} \;, \label{eq: KnudsenNumber} \\
    L_T = \frac{T_e}{|\nabla T_e|}\;, \\
    \lambda_{ei} = \frac{16 \pi T_e^2 \epsilon_0^2}{n_eZ^*e^4 \ln\Lambda_{ei}} \; , \;Z^* = \frac{\langle Z^2 \rangle }{ \langle Z \rangle} \; , 
\end{gather}
where $\lambda_{\textrm{col}}$ is a characteristic collision length for thermal electrons and $L_T$ is a characteristic temperature scale length. We've assumed a single $\log \Lambda_{ei}$ for all ion-species and quasi-neutrality $n_e=\langle Z\rangle n_i$. In the usual definition of Knudsen number, $\text{Kn}_{ei}\equiv  \lambda_{ei}/L_T$, with $\lambda_{ei}$ being the thermal electron-ion mean free path. The major assumption in a Chapman-Enskog derivation for fluid transport models is that one can expand the distribution function in increasing orders of Kn, which is assumed small.\cite{Spitzer, Braginskii}. Different terms in the kinetic equation apply at differing spatial and temporal orders with a closure relation needed to relate some highest order term to lower order terms.\cite{CartesianExpansion}. Small Kn means the plasma is in a highly collisional regime such that local thermodynamic equilibrium is nearly achieved. 

The reason Knudsen numbers exceeding 1\%, as opposed to exceeding order unity, can introduce nonlocal effects in heat transport is fast moving electrons ($v=3.6v_{th}$) contribute more to the heat flow compared to slow moving electrons\cite{Spitzer}. A more accurate Kn number should be weighted by the relative contribution of different electrons to the process of interest. However, because this is rarely done, Kn is instead based on thermal electrons at $v_{th}$. For heat transport, particles with $v\approx3.6 v_{th}$ contribute the most energy to the heat flow and because $\lambda_{ei}(v)\propto v^4$, the proper ratio for Knudsen number to indicate a breakdown of local fluid models is $\textrm{Kn}_{ei} > 1/(3.6)^4 \approx 0.01$.\cite{Bell1981} 

While a Knudsen number defined by $\lambda_{ei}$ is a common figure of merit, we find a more useful characteristic collision length is the delocalization length\cite{EpperleinShort, Davies} given by
\begin{gather}
	\lambda_e^* = \frac{4\pi \epsilon_0^2T_e^2 }{n_e e^4(Z^*\Phi \ln{\Lambda_{ei}}\ln{\Lambda_{ee}})^{1/2}} \label{eq: delocLength} \;\;\;\; , \\
	\Phi = \frac{Z^* + 4.2}{Z^* + 0.24} \label{eq: eeCorection} \;\;\; ,
\end{gather}
where $\Phi$ is a numerical correction to account for electron-electron self collisions: $\Phi=1$ for a Lorentz gas with just electron-ion collisions. Kn$_d \equiv \lambda_e^* / L_T$ is used for the remainder of this paper (see appendix).

A simple geometry of the problem is shown in fig. (\ref{fig: simpleOutline}). In MagLIF gaspipes, we find the largest impact from nonlocal effects occurs in the radial direction due to the much longer laser attenuation length in $z$ compared to $r$. This causes larger temperature gradients and thus more nonlocal conditions. 

This paper investigates local and nonlocal heat transport models for simulating the MagLIF preheat experiments on NIF. Although we primarily look at the NIF gaspipe experiments due to their simplicity in physics (no atomic physics, relatively little shocks, etc.), these efforts are meant to illuminate a general characterization of when local versus nonlocal versus kinetic models are needed in a variety of plasma systems. This paper is organized as follows. Section 2 is a description of the Vlasov-Fokker-Planck (VFP) equation and how the VFP code K2\cite{K2ref} extracts kinetic heat flows. Section 3 describes how reduced models are formed from the kinetic equation. Section 4 describes our fluid simulations using the radiation-hydrodynamic code HYDRA\cite{HydraRef} to create plasma profiles to initialize the VFP code K2. Section 5 shows primary results comparing radial heat flow from the three models: kinetic calculations from K2, a nonlocal heat flow developed by Schurtz et al. \cite{SNB}, and the local fluid model. Section 6 discusses the effects of a suppressed radial heat flow on laser propagation. In the appendix, we show the delocalization Knudsen number figure of merit and that Kn$_d$ exceeding 0.3\% can impact the heat flow propagation in these gaspipes. We have shown for unmagnetized gaspipe simulations, Kn$_d$ gives a metric for when nonlocality impacts the hydrodynamic evolution. Using Kn$_d$ in simulations with a magnetic field, external or self-generated, would be an interesting extension to the methodology as the problem becomes more complex with magnetic field scale lengths.

\section{Kinetic heat flow and VFP code K2}
We describe transport phenomena for our weakly-coupled plasma by solving the VFP equation:
\begin{gather}
    \pdv{f_e}{t} + \vb{v}\cdot\nabla f_e - \frac{e}{m_e} (\vb{E} + \vb{v} \times \vb{B}) \cdot \pdv{f_e}{\vb{v}} = {C}_{ee}(f_e) + C_{ei}(f_e) \label{eq: VFP}
\end{gather}
where $f_e$ is the electron distribution function, $\vb{E}$ and $\vb{B}$ are the electric field and magnetic field, and $C_{ee}$ and $C_{ei}$ are the collision operators for self-collisions and electron-ion collisions.
For \neop{} we do not treat the species separately and instead use $Z^*$ defined by eq. (\ref{eq: eeCorection}) to approximate multi-ion species effects.   
This equation is solved using the VFP code K2 by expanding $f_e$ in spherical harmonics in momentum space following the KALOS formalism\cite{Kalos,K2ref} given as
\begin{gather}
    f_e(\vb{r},\vb{v},t) = \sum_{n=0}^{n_{max}}\sum_{m=-n}^{n} f_n^m(\vb{r},v,t)P_n^{|m|} (\cos{\theta})e^{im\phi} \label{eq: feExpand} \;\;\; ,
\end{gather}
where $n$ and $m$ are the orders of spherical harmonics and $\theta$ and $\phi$ are the velocity-space polar (with respect to $v_x$ axis) and azimuthal angles.  An advantage of using spherical harmonics to study collisionally dominated plasmas is that higher orders of $f_n^m$ decay due to electron-ion collisions by $\exp(-\nu t)$, $\nu=\order{n(n+1/2)}$ so results converge with fairly few harmonics. However, this also means that the total distribution function does not automatically maintain positivity as truncation of the expansion can lead to $f_e$ becoming negative. We monitor how negative $f_e$ becomes, and for our simulations shown this effect does not cause a numerical instability. 

K2 currently solves $C_{ei}$ with stationary ions as a pitch angle scattering operator using Epperlein and Short's\cite{EpperleinShort} numerical correction factor $\Phi$ for arbitrary $Z$. $C_{ei}$ is given by
\begin{gather}
C_{e i}\left(f_{e}\right)=\frac{n_i \Gamma_{ei}}{v^{3}} \Phi \mathcal{L}\left(f_{e}\right)\;\;, \\ \mathcal{L}=\frac{1}{2}\left[\frac{\partial}{\partial \mu}\left(1-\mu^{2}\right) \frac{\partial}{\partial \mu}+\frac{1}{1-\mu^{2}} \frac{\partial^{2}}{\partial \phi^{2}}\right] \;\;, \\
\Gamma_{ei} = \frac{e^4}{4\pi\epsilon_0^2m_e^2} \left(Z^*\right)^2 \ln{\Lambda_{ei}} .
\end{gather}
Where $\mu=\cos{\theta}$. The isotropic self-collisions follow the form laid out in Tzoufras \textit{et al.} \cite{TzoufrasOSHUN} and reproduced here as
\begin{gather}
\frac{C_{ee0}}{\Gamma_{ee}} = \frac{1}{v^2} \pdv{v} \left[\mathcal{C} (f_0) + \mathcal{D}(f_0) \pdv{f_0}{v}\right] \;\;,\\
\mathcal{C}(\vb{r},v,t) = 4 \pi \int_0^v f_0 (\vb{r},u,t)u^2 \dd u \;\; , \\
\mathcal{D}(\vb{r},v,t) = \frac{4\pi}{v} \int_0^v u^2 \int_u^\infty f_0(\vb{r},v',t)v' \; \dd v' \dd u \; , \\
\Gamma_{ee} = \frac{e^4}{4\pi\epsilon_0^2m_e^2} \ln\Lambda_{ee} \;\;\; ,
\end{gather}
where $\mathcal{C}$ and $\mathcal{D}$ are the Rosenbluth potentials\cite{Rosenbluth}. Although K2 has an implementation for anisotropic self-collisions, they were turned off throughout our work and only used once to verify the anisotropic self-collisions did not significantly impact our results versus using $\Phi$ from Eq. (\ref{eq: eeCorection}).  A derivation of the anisotropic operator used can be found in equation (41) of Tzoufras \textit{et al.} \cite{TzoufrasOSHUN}. The Coulomb logarithms $\ln{\Lambda}$ for electron-electron and electron-ion scattering have the form $\Lambda = (b_{max} / b_{min})$. For simplicity we assume the same Coulomb logarithms for both e-e and e-i collisions. We also assume for \neop{} that a single $\ln{\Lambda}$ formulated with $Z^*$ is sufficient compared to using separate Coulomb logarithms for each species. $b_{max}$ (the maximum impact parameter) is the smaller of the Debeye length or the ion sphere radius and $b_{min}$ (the minimum impact parameter) is the larger of the de Broglie wavelength $\lambda_b = \hbar / 2 m_e v_{th}$ and the distance of closest approach $\rho = Z^* e^2 / (4\pi \epsilon_0 m_e v_{th}^2)$. \cite{Huba2004NRLPF} 

The electric and magnetic fields are handled by solving the full Maxwell equations with a slight modification on Ampère's Law to allow larger time steps for faster computation given by
\begin{gather}
	\pdv{\vb{B}}{t} = -\nabla \times \vb{E} \;\; , \;\; \pdv{\vb{E}}{t} = \alpha\left(c^2\nabla \times \vb{B} - \frac{\vb{J}}{\epsilon_0}\right) \;\;\;,
\end{gather}
where $\alpha$ is a small dimensionless numerical factor that alters the electric field response due to plasma dynamics. This factor yields a dispersion relation for EM waves  $\omega^2 = \alpha(\omega_p^2 + k^2c^2)$ and plasma waves $\omega^2 = \alpha\omega_p^2$, which lengthens the plasma wave period as $1/\sqrt{\alpha}$. 
Therefore, this allows us to choose a larger timestep $\Delta t \lesssim 1/\omega_p \sqrt \alpha$ and avoid numerical instability due to under-resolving wave motion. 
The choice of $\alpha$ is left to the user but the most important physics for our problem is to ensure that the effective plasma period is much smaller than the simulation time. 
While this condition provides a preliminary choice for $\alpha$, ultimately, we ran several 1D convergence tests for our problems to ensure the transport was not affected by the choice of $\alpha$. We find $\alpha = 1/6400$ to be a good choice for our simulations.

Once $f_e$ has been computed by solving Eq. (\ref{eq: VFP}), the heat flow can be calculated using
\begin{gather}
\vb{Q} = \frac{m_e}{2}\int f_e v^2 \vb{v}\dd^3 \vb{v} \;\;\; .
\end{gather}
The steepest gradients with the shortest scale length in MagLIF simulations occur in the radial direction so we expect the highest nonlocality to occur in the radial direction. While our primary comparisons to classical theory will be 1D comparisons, gradients in the other dimensions can affect the heat flow. 2D simulations are left for future work due to their computational costs.

\section{Local vs. Reduced Non-Local Heat flow Models}
Solving the VFP equation is computationally expensive, so modeling transport phenomena requires simplified fluid models that can be computed quickly to evolve plasma parameters. The most famous of these models are the classical Spitzer-Härm unmagnetized heat flow model \cite{Spitzer} and the Braginskii transport model for magnetized plasmas \cite{Braginskii}. Subsequent publications have corrected small inaccuracies\cite{EpperleinHaines} in the models while retaining their basic character. These fluid computations of transport phenomena rely on assuming a Chapman-Enskog expansion in a small parameter typically the Knudsen number. This means that a primary assumption in these theories is that the scale length of the problem is much longer than the electron-ion mean free path. With this assumption, one can simplify the heat flow calculation to be simply dependent on a thermal conductivity and a temperature gradient: $\vb{Q} = -\kappa \cdot \nabla T_e$. While these theories work well with gentle temperature gradients, this condition is generally not satisfied in ICF experiments.\cite{EpperleinShort,Bell1985}

When the local approximation condition (Kn $\ll 1$) breaks down, one needs to consider kinetic effects. A common kinetic effect is the reduction of heat flow in the largest heat flow regions before the heat front\cite{Bell1985}. Nonlocality smears out the superthermal electron deposition making the heat flow smaller in some regions and larger in others. In laser-heated plasmas, regions with high electron temperature, but smaller temperature gradients, experience a reduced heat flow. Outside of the heated region is colder plasma with larger temperature gradients which experience enhanced heat flow (also known as the preheat). We are mostly interested in modeling the laser-heated region so capturing the heat flow reduction is more important. To try and capture this heat flow reduction, codes commonly implement an artificial flux limit $f_L$ typically in the range of 3\% to 15\% \cite{Goncharov2005,Goncharov2006,Rosen2011}: $\vb{Q}_{fL} = min[\vb{Q}_{\textrm{local}},f_L * n_e T_e v_{th}]$.
Alternative models to the flux-limiter include a harmonic flux-limiter\cite{Davies,Walsh2024} $\vb{Q}_{fH} = \vb{Q}_{\textrm{local}} / (1 + |\lambda_e^* / f_H L_T |)$. (see sec. \ref{sec: K2sims})
However, $f_L$ and $f_H$ are still determined empirically and vary between problems.

In attempts to capture more nonlocal effects without the numerical cost of a full VFP simulation, several "reduced" nonlocal models have been proposed. A common idea is to somehow incorporate energetic electrons from surrounding regions into the heat flow calculation. This is mathematically done with a convolution between a local theory of heat flow and a kernel $W$ representing the probability an electron deposits its energy from one region to another\cite{Bell1983, LMV, CMG, SNB}. 
A popular nonlocal model is one proposed by Schurtz, Nicolaï, and Busquet \cite{SNB} known as the SNB model. The SNB model used in our analysis is given by
\begin{gather}
    \left(\frac{2}{\lambda_{g,ee}(r)} - \nabla \cdot \frac{\lambda_{g,ei}^E(r)}{3} \nabla \right) H_g(r) = -\nabla \cdot \vb{U}_g(r) \label{eq: SNBmain} \;\;\; ,\\ 
    \vb{U}_g(r) = \frac{1}{24} \int_{\beta_{g-1}}^{\beta_g} \beta^4 e^{-\beta} \dd \beta \vb{Q}_{\textrm{local}}(r) \;\;\; 
\end{gather}
where $g$ denotes energy group, and we have incorporated separate mean free paths $\lambda_{g,ee}$ and $\lambda_{g,ei}^E$  following Brodrick's improvements to the original SNB model\cite{BrodrickCorrection,Ma2022}. $\beta \equiv \frac{1}{2} m_e v^2 / T_e$ and $\vb{U}_g$ is integrated over energy group $g$. The nonlocal $\vb{Q}$ is then:
\begin{gather}
    \vb{Q}(r) = \vb{Q}_{\textrm{local}}(r) - \sum_g \frac{\lambda_{g,ei}^E(r)}{3} \nabla H_g(r) \;\;\; .
\end{gather}
The SNB model also incorporates a mean free path modification limited by the contribution due to the local (Spitzer-Härm) electric field defined as
\begin{gather}
    \frac{1}{\lambda_{g,ei}^E} = \frac{2}{\sqrt{Z+1}}\frac{1}{\lambda_{g,ei}} + \frac{\abs{e \vb{E}^{SH}}}{T_e\beta_g} \;\;\; .
\end{gather}
This circumvents convergence issues at low collisionality and improves the modeling near plasma boundaries \cite{SNB}.
SNB has been implemented in a variety of ICF codes including HYDRA, LILAC, and DUED, and has been benchmarked in a variety of test problems \cite{K2ref, BrodrickCorrection,Marocchino}. The SNB model is a popular choice as it is designed to be solved in parallel over energy groups through multi-group diffusion algorithms.

\begin{figure*}[ht]
    \centering
    \begin{subfigure}{0.3\textwidth} 
        \centering
        \includegraphics[width=\textwidth]{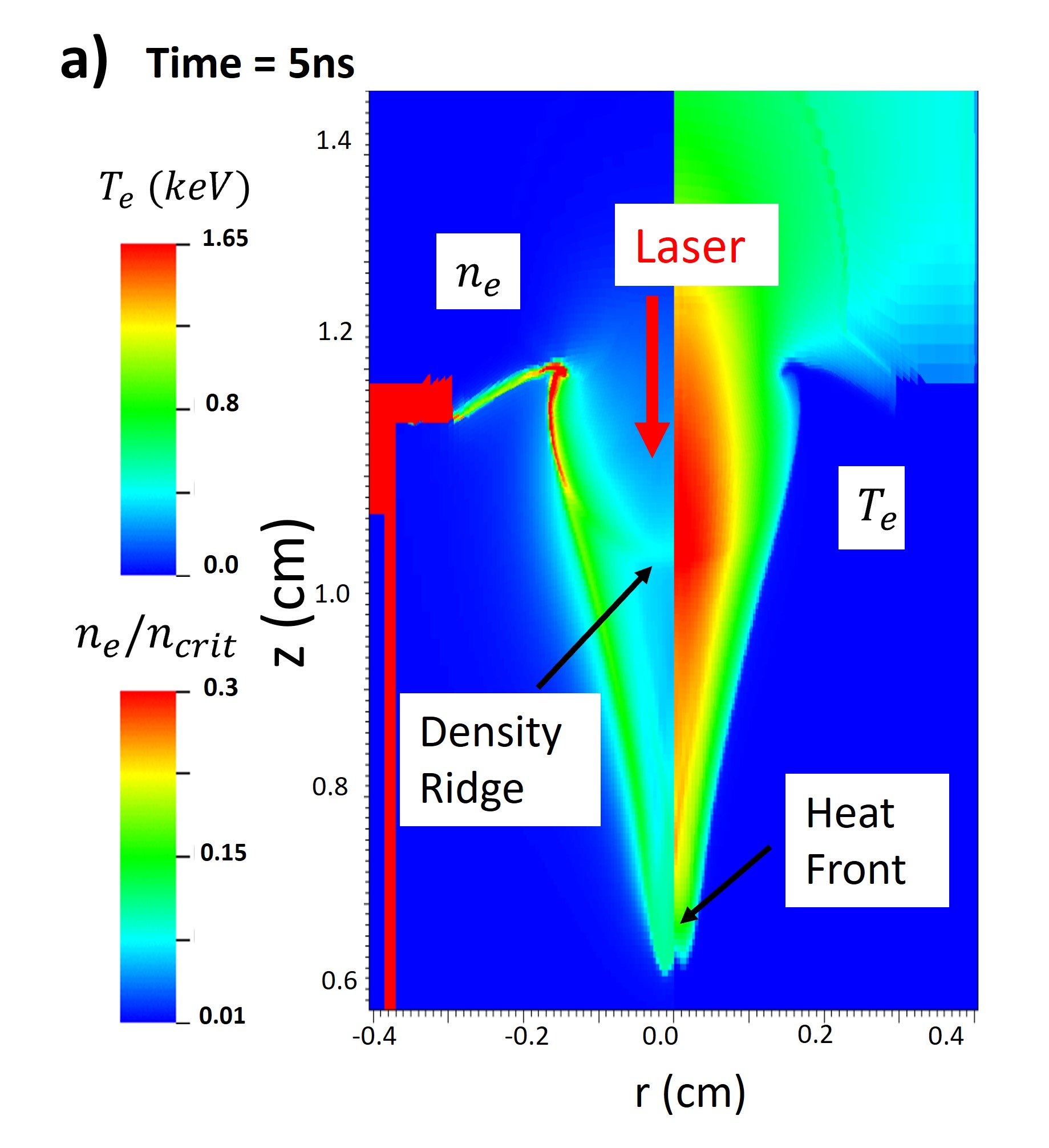}
        \phantomsubcaption
        \label{fig: C5H12early}
    \end{subfigure} 
    \hfill
    \begin{subfigure}{0.3\textwidth}
        \centering
        \includegraphics[width=\textwidth]{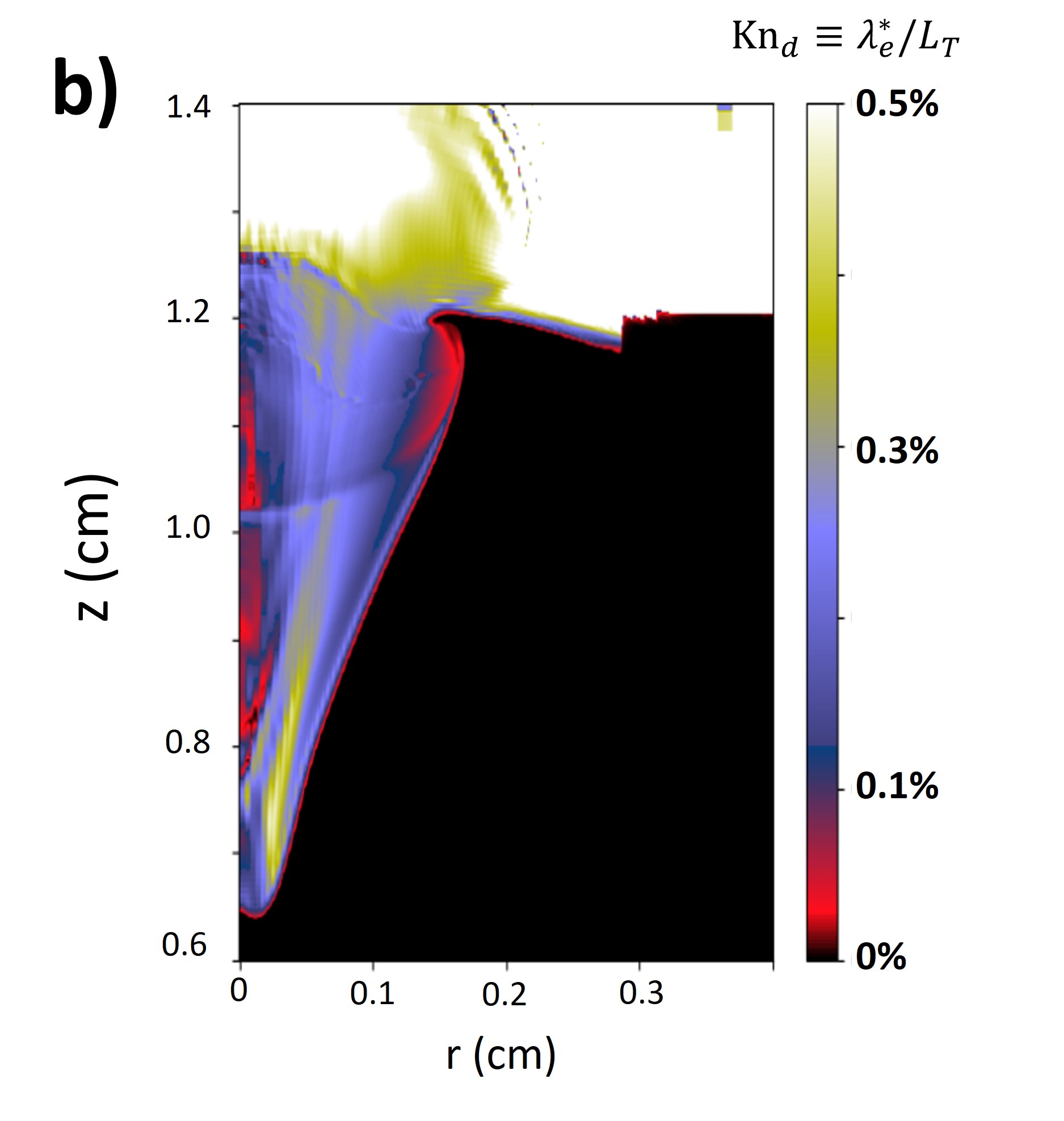} 
        \phantomsubcaption
        \label{fig: C5H12kn}
    \end{subfigure} 
    \hfill
    \begin{subfigure}{0.35\textwidth}
        \centering
        \includegraphics[width=\textwidth]{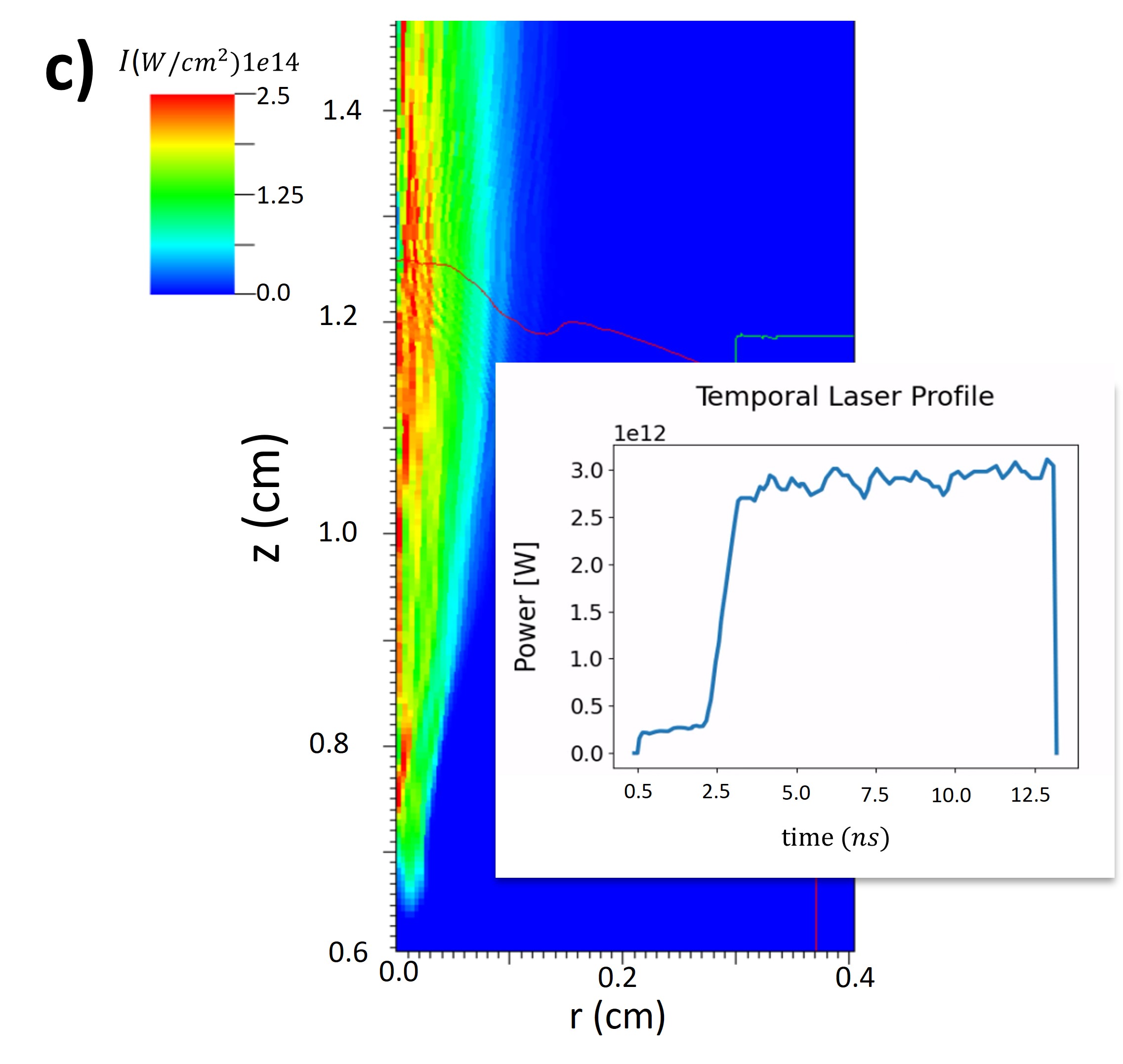}
        \phantomsubcaption
        \label{fig: lspatial}
    \end{subfigure} 
    \hfill
\caption{
3 mg/cc neopentane (\neop{}) gas fill HYDRA simulation at $t= 5$ ns: \textbf{\subref{fig: C5H12early})} electron density (left) and electron temperature (right). $n_{crit} \approx 9\times10^{21}$ cm$^{-3}$ is the critical density associated with a 351-nm wavelength laser. \textbf{\subref{fig: C5H12kn})} Knudsen number $\equiv \lambda_{e}^*/L_T$, exceeds 0.3\% where local heat conduction becomes invalid. \textbf{\subref{fig: lspatial})} Laser intensity and total power history.}
\label{fig: firstHydraSim}
\end{figure*}

Regardless of the method of how heat flow $\vb{Q}$ is computed, the electron temperature $T_e$ is evolved by the energy equation
\begin{gather}
    \frac{3}{2}n_e\pdv{T_e}{t} = -\nabla \cdot \vb{Q} + ... \;\;\; ,
\end{gather}
where ... stands for additional terms such as radiation or laser absorption. Radiative transport is sub-dominant in these low-Z gas-filled pipes so our rad-hydro simulations do not include it. Laser absorption is calculated via a standard inverse-bremsstrahlung model.

\section{Fluid Gaspipe Modeling with HYDRA}
We investigate the impact of kinetic effects on heat transport in MagLIF gaspipes by first using the rad-hydro code HYDRA to generate relevant conditions for the kinetic code K2 because a kinetic simulation on nanosecond time scales is too computationally expensive. The gaspipe is 1 cm long with an outer radius of 3.85 mm with 150-$\mu$m thick parylene walls. A 351-nm laser with a focused beam radius of approximately 0.6 mm enters through a 1-$\mu$m thick kapton window and travels through the gaspipe delivering laser energy through inverse-bremsstrahlung. HYDRA treats laser propagation through a geometric-optics, ray-tracing model. Neopentane gas \neop{} ($\langle Z^2\rangle/\langle Z\rangle \approx 4.5$) and deuterium gas ($\langle Z^2\rangle /\langle Z\rangle=1$) fills are studied with densities ranging from 3 mg/cc - 7.5 mg/cc to replicate current experiments.\cite{MagLIFexp} 
Neopentane has similar hydrodynamic structures to deuterium but can be fielded at a much lower gas fill pressure than deuterium for the same desired electron density. The comparable shot with deuterium requires cryogenic cooling and thicker windows making it more difficult to accomplish. 
For 3 mg/cc neopentane and 5 mg/cc deuterium this results in a fully ionized electron density of $\approx 11.5\% n_{crit}$ where $n_{crit}$ is the critical electron density for the laser wavelength. Separate electron and ion temperatures are used. The main physics included in our simulation is hydrodynamic motion, thermal conduction, laser propagation and laser absorption. Radiation was determined to be less dominant than these effects for the low-Z gaspipes.  Magneto-hydrodynamic effects were also not considered for simplicity. 

Hydra uses an arbitrary Lagrangian-Eulerian mesh that evolves with plasma flow.  We interpolate plasma conditions from this mesh onto a uniform grid for K2 simulations. We focus on the 1D radial spatial domains between 0.15 cm - 0.2 cm, because outside of this region the radial heat flow is much smaller. In $z$, we focus on spatial domains between 0.1 cm - 1 cm to avoid a hydrodynamic feature generated by burndown of the laser entrance window. Fig. \ref{fig: C5H12early} shows an example of a HYDRA simulation at one of the time slices we investigate. 
We then compute Kn$_d$ (example Fig. \ref{fig: C5H12kn}) to evaluate regions where nonlocality might be important as a full VFP simulation on the full domain is impractical and unnecessary. We find Kn$_d \ge 0.3\%$ is a good threshold for the onset of nonlocal effects, where local electron conduction theory overpredicts the kinetic heat flow by a factor of $\approx$1.2 (see appendix). For example, a time late into the gaspipe ($t=8.5$ ns for 3 mg/cc neopentane and $t=9.75$ ns for 5 mg/cc), where the laser has reached the end of the gaspipe, there are peak Kn$_d = 0.55\%$ for neopentane and  Kn$_d = 0.32\%$ for deuterium. While neopentane has a higher average $Z^*$ than deuterium, shortening the delocalization length, neopentane gas experiences more laser absorption, increasing electron temperature. The increased electron temperature induces larger temperature gradients, causing a higher Kn$_d$  in neopentane than deuterium. 

Three different time slices are studied corresponding to roughly when the laser has propagated 33\%, 50\% , and 90\% through the gaspipe to see when nonlocal effects may develop.  We avoid investigating conditions too early or too late into the laser propagation as burn down of the entrance and exit windows is beyond the scope of this paper. 

\section{K2 Vlasov-Fokker-Planck Simulations on HYDRA Output} \label{sec: K2sims}
\begin{figure}
\begin{center}
\includegraphics[scale=0.5]{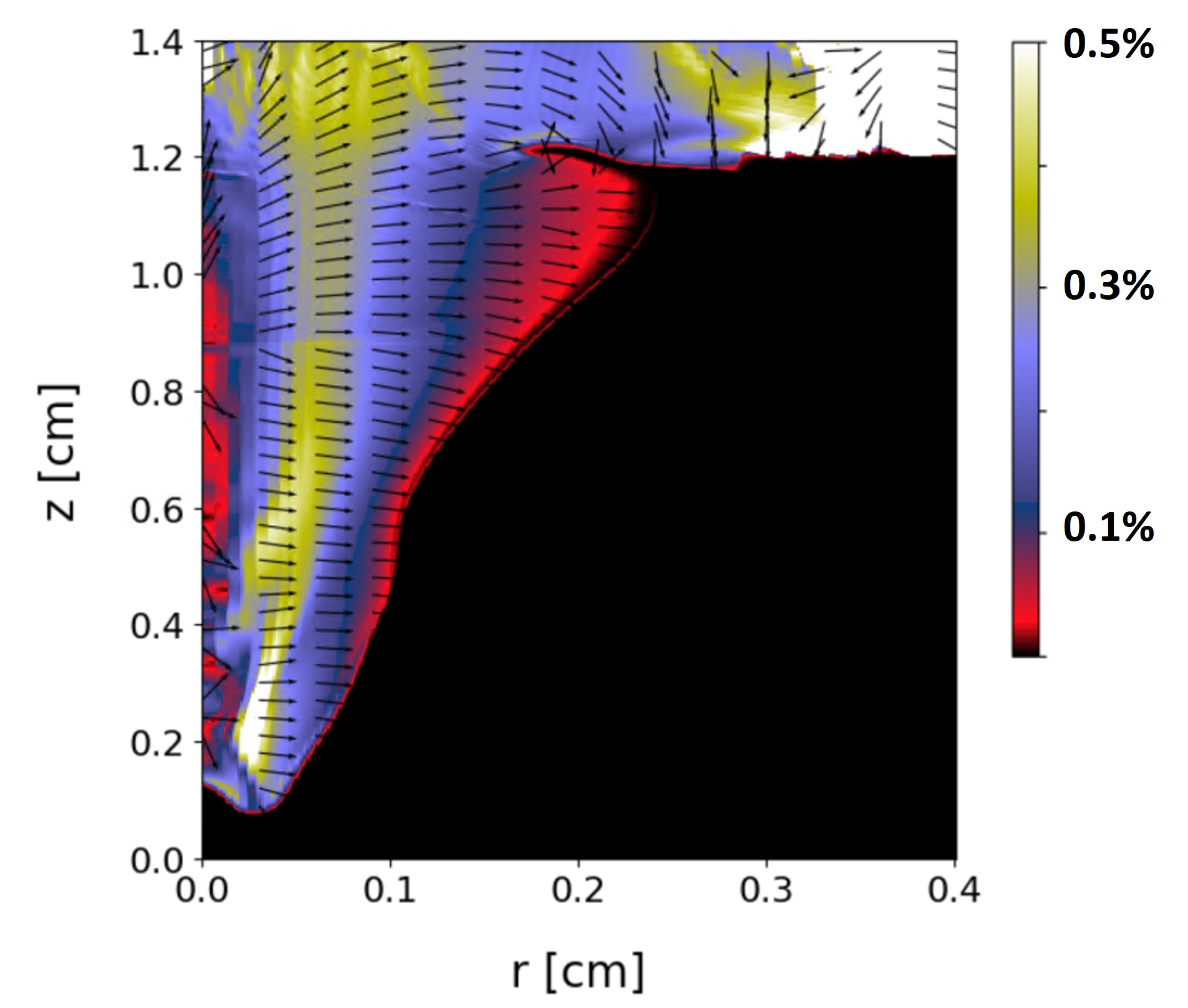} 
\end{center}
\caption{Kn$_d$ plot of the HYDRA simulation shown in Fig. \ref{fig: firstHydraSim} as the laser reaches the end of the gaspipe at $t\approx10$ ns. The vector arrows indicate the direction of heat flow $\vb{Q}_{\textrm{local}} / |\vb{Q}_{\textrm{local}}|$. Heat flow beyond $r=0$ is primarily in the radial direction, making it more interesting to investigate nonlocality in radial rather than axial heat flow.} 
\label{fig: vFieldKn}
\end{figure}

This paper primarily focuses on radial heat flow $\vb{Q}_{\perp}$ because the axial heat flow is mostly local. The gradients along $z$ are not drastic making $L_T$ much longer than $\lambda_e^*$. This is shown in figure \ref{fig: vFieldKn} where the vector arrows are the direction of heat flow $\vb{Q}_{\textrm{local}} / |\vb{Q}_{\textrm{local}}|$. Beyond $r=0$, where the laser propagates through the gaspipe with more complicated dynamics, $\vb{Q}$ points strongly along $\vb{r}$. This is verified by our 1D axial kinetic calculations which show that the local model is sufficient for capturing the axial heat flow well beyond the density ridge in Fig. \ref{fig: C5H12early}. 

Two $z$ slices, defined by the heat front and the density ridge, were chosen to initialize 1D radial profiles for K2. The heat front is defined as the smallest $z$ (furthest $z$ from the entrance window) where $T_e > 1$  keV. The density ridge is defined as the largest $z$ (closest $z$ from the entrance window) along $r=0$ where $n_e > 8.5 \times 10^{20}$  cc$^{-1} \approx 9.5\% n_{crit}$. 
1D radial heat flows are measured at the heat front and halfway in between the heat front and the density ridge. We generally find that these areas consistently include the peak Kn$_d$ and are good candidates for nonlocal behavior. 

\begin{figure}
\raggedright
\includegraphics[scale=0.6]{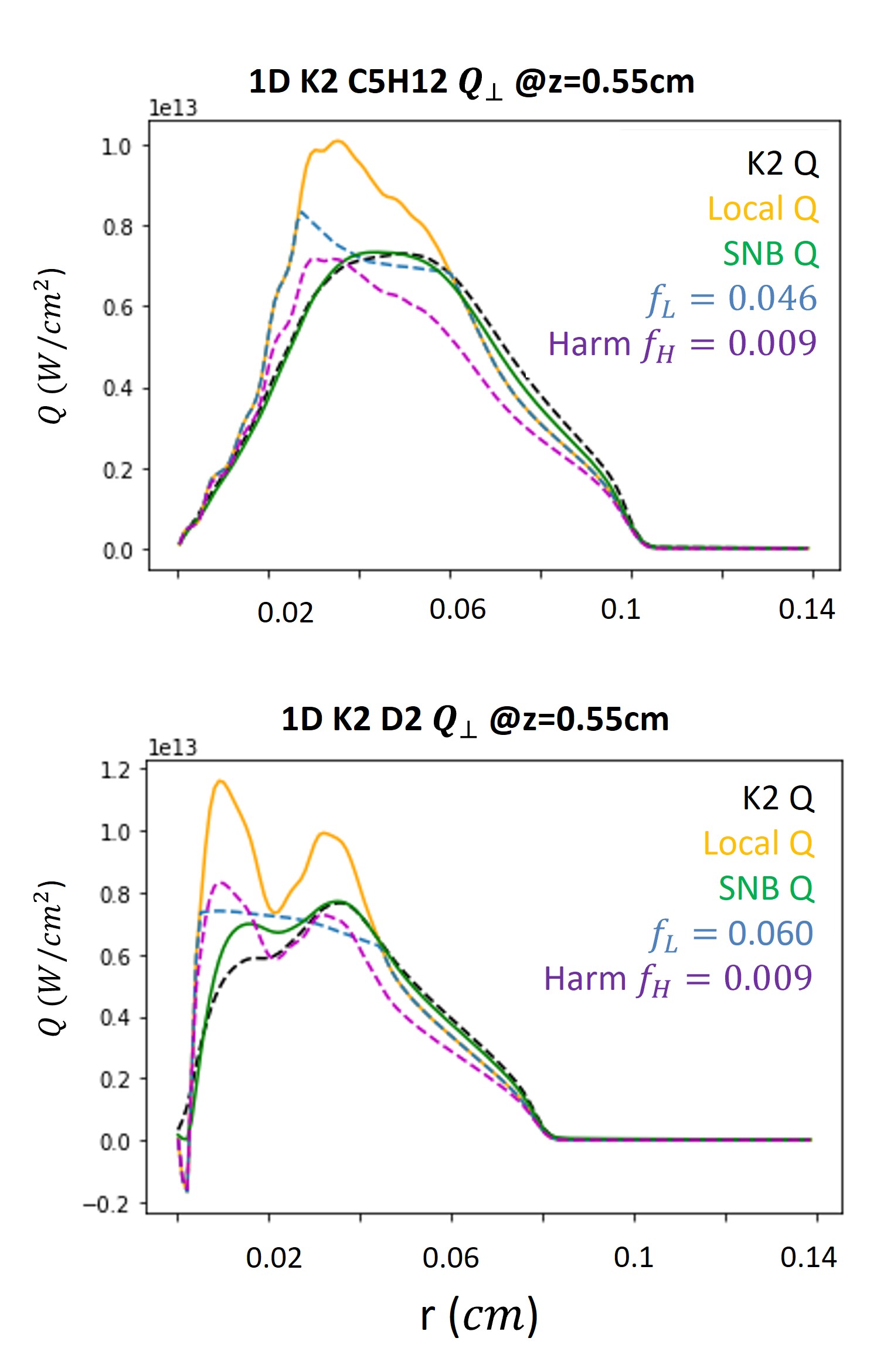} 
\caption{1D K2 outputs, initialized with HYDRA generated plasma conditions (fig. \ref{fig: firstHydraSim}), comparing radial heat flows $Q_\perp$ between a local theory (Epperlein-Haines), kinetic calculations produced by K2, and an SNB model computed heat flow. Blue dashed lines are the common flux-limited model $\vb{Q}_{fL} = \min[\vb{Q}_{\textrm{local}},f_L * n_e T_e v_{th}]$ and purple dashed lines the harmonic flux model $\vb{Q}_{fH} = \vb{Q}_{\textrm{local}} / (1 + |\lambda_e^* / f_H L_T |)$. For peak Kn$_d = 0.55\%$ we observe that local theory over predicts kinetic theory by 1.6x in central regions $r=0.04$ cm for \neop{}.}
\label{fig: K2C5H12} 

\end{figure}

\begin{figure}
\centering
\begin{subfigure}{0.45\textwidth} 
    \centering
    \includegraphics[width=\textwidth]{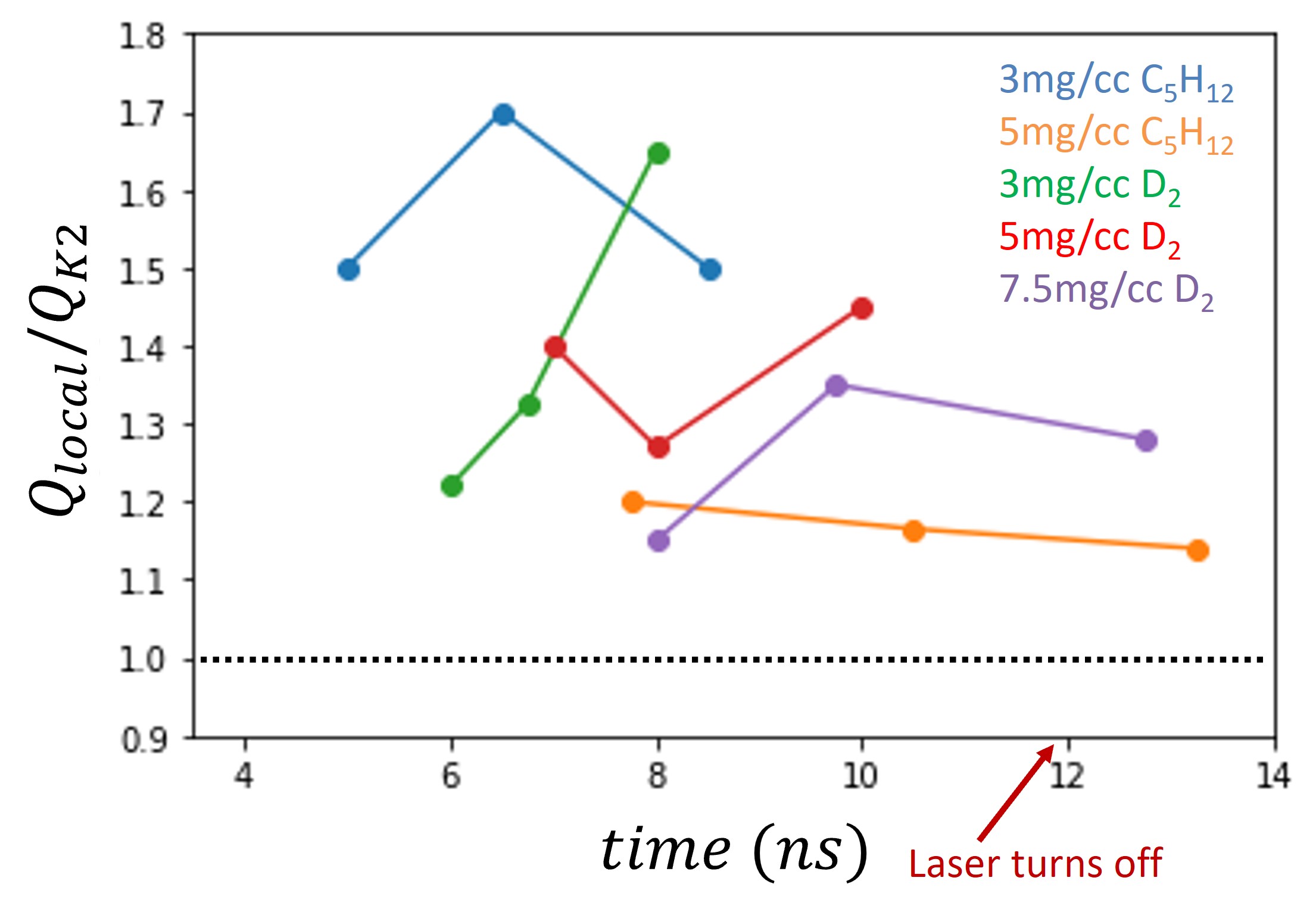}
    \phantomsubcaption
    \label{fig: GaspipeParamSearch}
\end{subfigure}
\hfill
\begin{subfigure}{0.45\textwidth} 
    \centering
    \includegraphics[width=\textwidth]{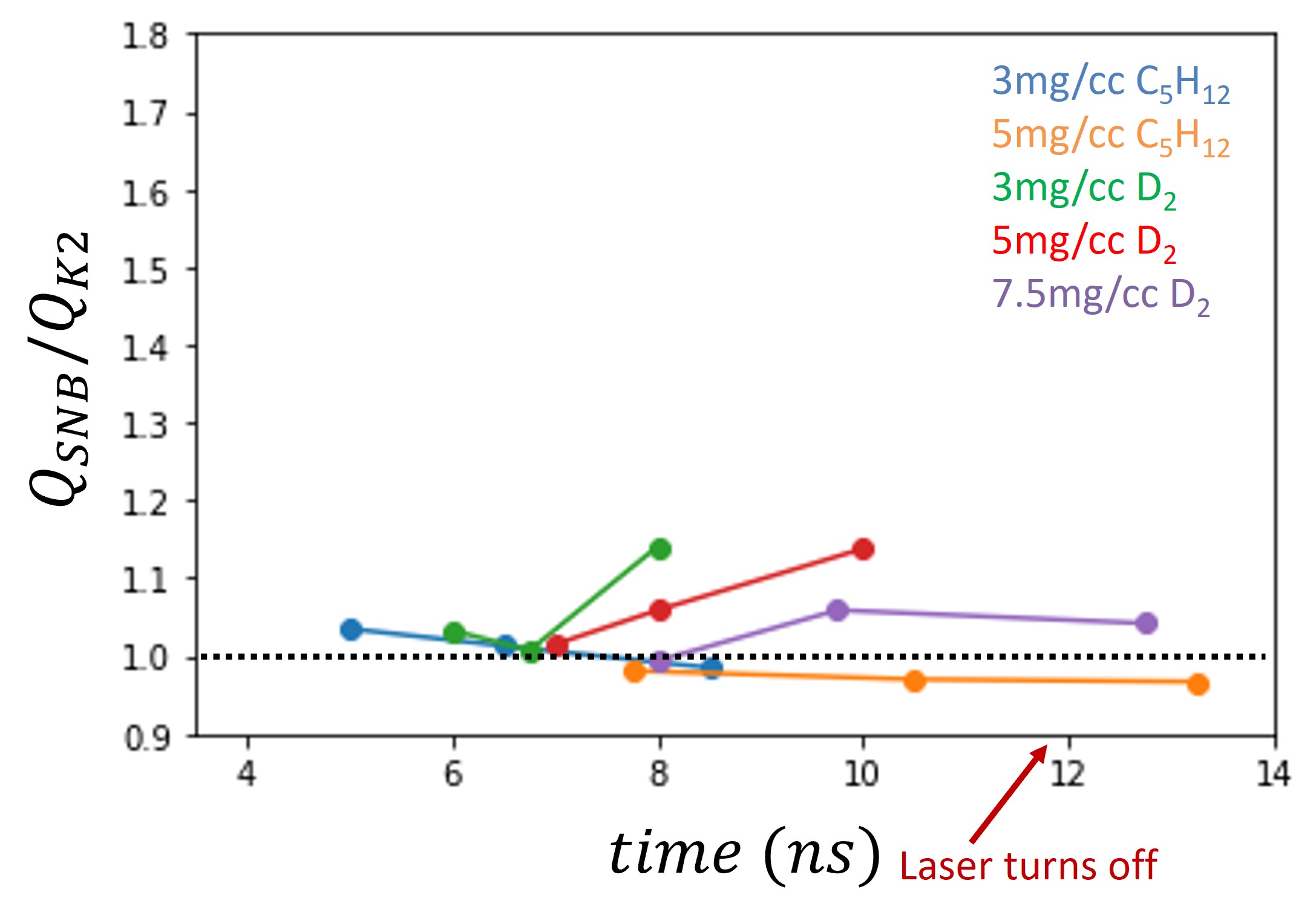}
    \phantomsubcaption
    \label{fig: GaspipeSNB}
\end{subfigure}
\hfill
\caption{Top: Ratio of the local heat flow $|\vb{Q}_{\textrm{local}}$|/$|\vb{Q}_{\textrm{K2}}|$ over $r=$max($\vb{Q}_{\textrm{local}}$) $\pm 50  \mu$m are shown. Solid points are times of early, middle, and late laser propagation into the gaspipe. The local model of electron conduction consistently over predicts the kinetic heat flow over the full hydrodynamic evolution. Bottom: Ratio $|\vb{Q}_{\textrm{SNB}}|$/$|\vb{Q}_{\textrm{K2}}|$ is close to 1. We find that SNB heat flow captures the kinetic result extremely well only deviating by a maximum $\approx$ 10\%. Data points are observed after $t=4$ ns to avoid the ablation of the kapton window.}
\label{fig: GasParamSearchTotal}
\end{figure}

We measure $Q_{\perp,\textrm{local}}$ and $Q_{\perp, \textrm{K2}}$ after the heat flow reaches a quasi steady-state ($t\approx 25\tau_{ei}$) using distribution functions solved by K2 from the initialized plasma parameters. To keep plasma parameters constant during this time, we also apply heating and cooling operators to maintain $T_e$ as close to its initial state as possible. 
The ratios of $Q_{\perp, \textrm{local}} / Q_{\perp, \textrm{K2}}$ are then computed within the region $r=$max($Q_{\perp,\textrm{local}})\pm 50 \ \mu$m to account for the shape of nonlocal heat flow. An example of a K2 simulation measurement is shown in figure \ref{fig: K2C5H12}. Generally, the kinetic radial heat flow extends further in the radial direction, because higher energy particles deposit their energy further away from the axis due to their long mean free path. This results in a flattened version of local theory.
The ratio, $|\vb{Q}|/|\vb{Q}_{local}|$, was then averaged within the measured window. This analysis is repeated for different $z$ values. This provides an overview of the radial heat flow over the gaspipe at the corresponding times.
Fig. \ref{fig: GasParamSearchTotal} demonstrates that a local theory of electron conduction consistently over predicts the kinetic result by a factor of $1.1 - 1.7$.

While the local model deviates noticeably from K2, we find that the reduced nonlocal SNB model captured kinetic effects very well. We measure the ratio of the heat flow computed from an SNB model and the kinetic K2 heat flow for the same gaspipe parameters. As shown in Fig. \ref{fig: GasParamSearchTotal}, we find the SNB heat flow deviates from the kinetic result by a factor of 0.98 to 1.14. Additionally, we find the profile of the heat flow is captured very well by the SNB model. An example shown in Fig. \ref{fig: K2C5H12} demonstrates that SNB not only reproduces the kinetic heat flow values well, but also the preheat phenomenon where the heat front is shifted to a slightly larger radius than the local heat flow model. 

The flux-limited models in Fig. \ref{fig: K2C5H12} have $f_L$ and $f_H$ to best match the average kinetic heat flow in the central region between $0.02$ cm $\leq r \leq 0.06$ cm in neopentane and $0.015$ cm $\leq r \leq 0.04$ cm in deuterium. The traditional flux-limited model $\vb{Q}_{fL} = min[\vb{Q}_{\textrm{local}},f_L * n_e T_e v_{th}]$ can match the kinetic heat flow well but only when gradients in $T_e$ and $n_e$ are smooth. The harmonic flux-limited model, $\vb{Q}_{fH} = \vb{Q}_{\textrm{local}} / (1 + |\lambda_e^* / f_H L_T |)$, can match the kinetic heat flow adequately as well, but can also better match the shape of the kinetic heat flow. This model also has the added benefit of a more consistent $f_H$ than $f_L$ choice between the two gas fills. For the traditional model to match the kinetic heat flow, $f_L$ varies between $f_L=0.046$ for neopentane and $f_L=0.06$ for deuterium. The harmonic model uses a single $f_H=0.009$ to match both neopentane and deuterium. This may be because $\vb{Q}_{fH}$ uses $\lambda_e^*$ which better accounts for variations in $Z$ in Kn$_d$ (see appendix).

In summary, we observe that the local heat flow consistently over predicts the kinetic heat flow by a factor $\approx 1.1-1.7$ throughout the entire hydrodynamic evolution of the gaspipes. We also observed that the SNB model consistently captures kinetic effects in the radial heat flow throughout the gaspipe evolution. 

\section{Effects of Reduced $Q_{\perp}$ in Heat front Propagation}
The overall effect of a reduced radial heat flow $Q_{\perp}$ leaves the center of the gaspipe, near $r=0$, hotter. The local heat flow advects heat too quickly radially into the gaspipe compared to the kinetic result, reducing the overall electron temperature. Because the SNB model captures kinetic effects well, we used the SNB heat flow conduction model in HYDRA to simulate the radial heat flow more accurately.
For this, we used Brodrick's improvements\cite{BrodrickCorrection}. We find using 60 electron energy groups to solve the multi-group diffusion problem gives sufficient convergence.
\begin{figure} [h!]  
\centering 
\begin{subfigure}{0.45\textwidth}
    \centering
    \includegraphics[scale=0.5]{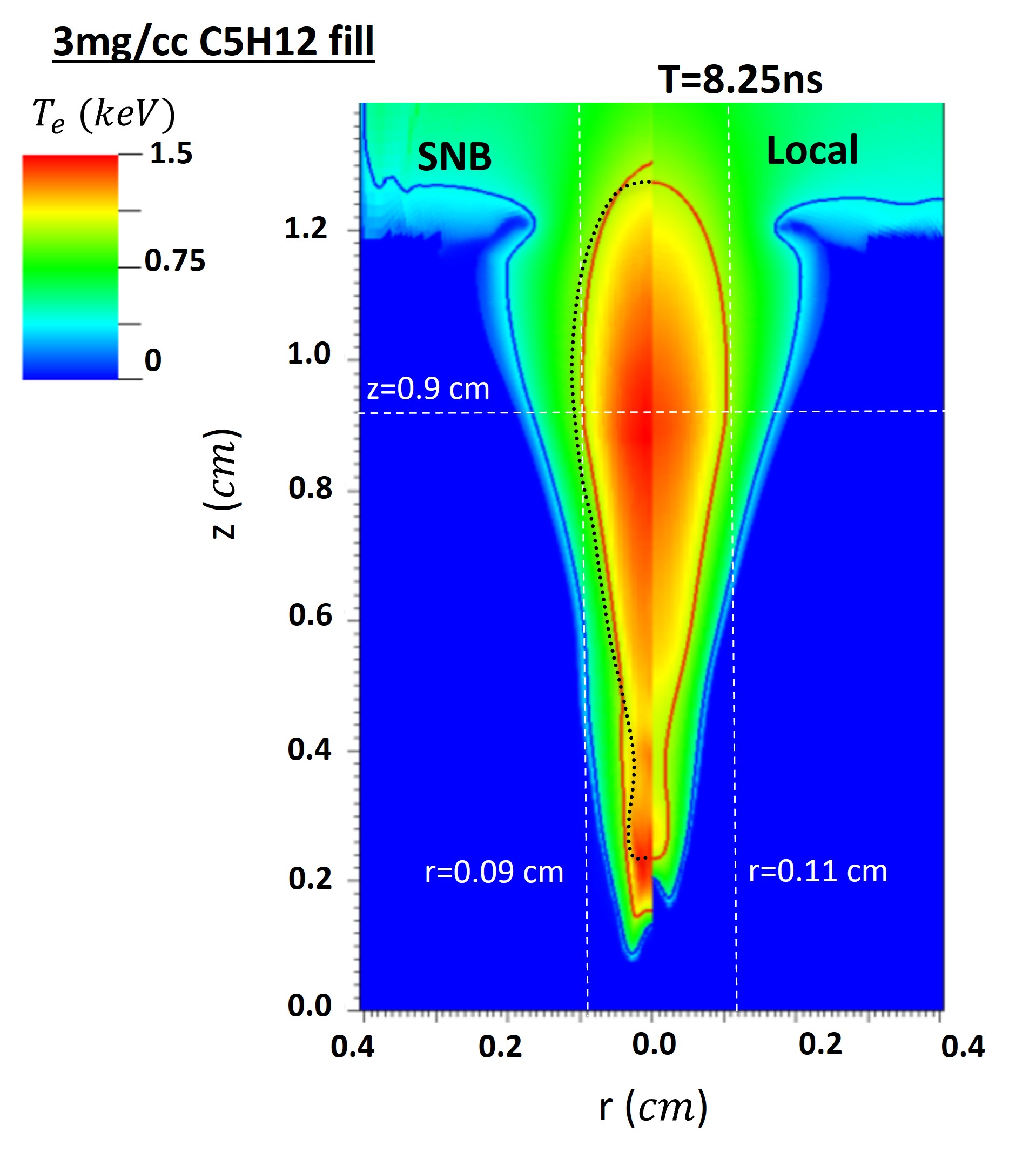} 
    \phantomsubcaption
\end{subfigure}
\hfill
\begin{subfigure}{0.45\textwidth}
    \centering
    \includegraphics[scale=0.4]{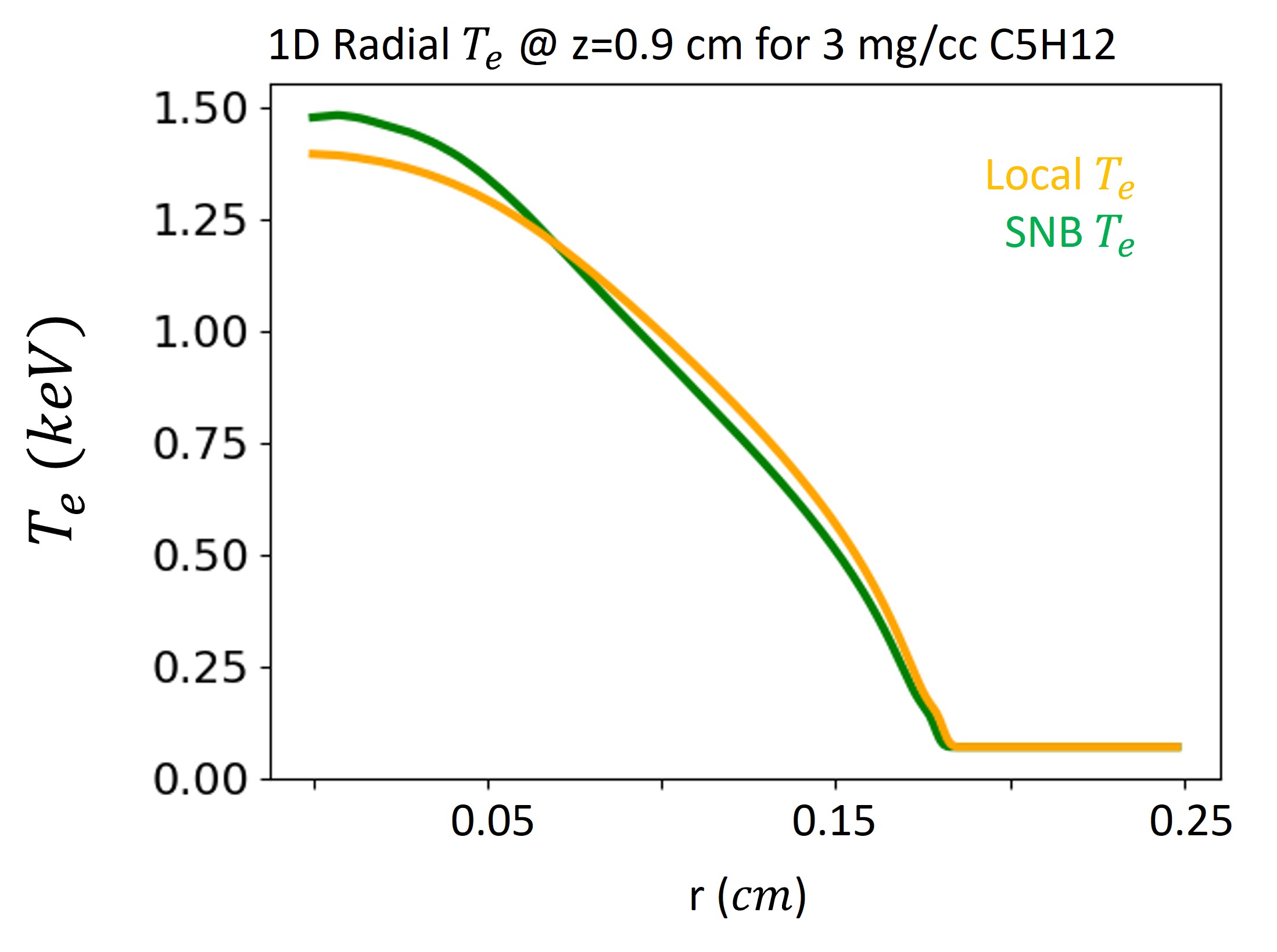}
    \phantomsubcaption
\end{subfigure}
\caption{\textbf{Top:} HYDRA simulation using SNB \cite{SNB} reduced nonlocal thermal conduction (left) and local (right) thermal conduction. 3 mg/cc of neopentane is $\approx$ 12\% $n_{e0} / n_{crit}$ for 351 nm wavelength light. Nonlocal physics increases the heat front propagation speed compared to local conduction. Contour lines (dark red = 1 keV, dark blue = 0.4 keV, black dotted = local 1 keV contour reflected onto the left side) are added to help visualize the increased temperature diffusion in the local thermal conduction model vs. the SNB model. Vertical white dotted grid lines correspond to the maximum length in \textit{r} for the 1 keV contour and the horizontal white dotted grid line corresponds to the largest differential between the SNB and local 1 keV contours. \textbf{Bottom:} 1D lineouts of electron temperature at $z=0.9$ cm.}
\label{fig: SNB_evolution}
\end{figure}

\begin{figure} [h!]

\centering 
\includegraphics[scale=0.5]{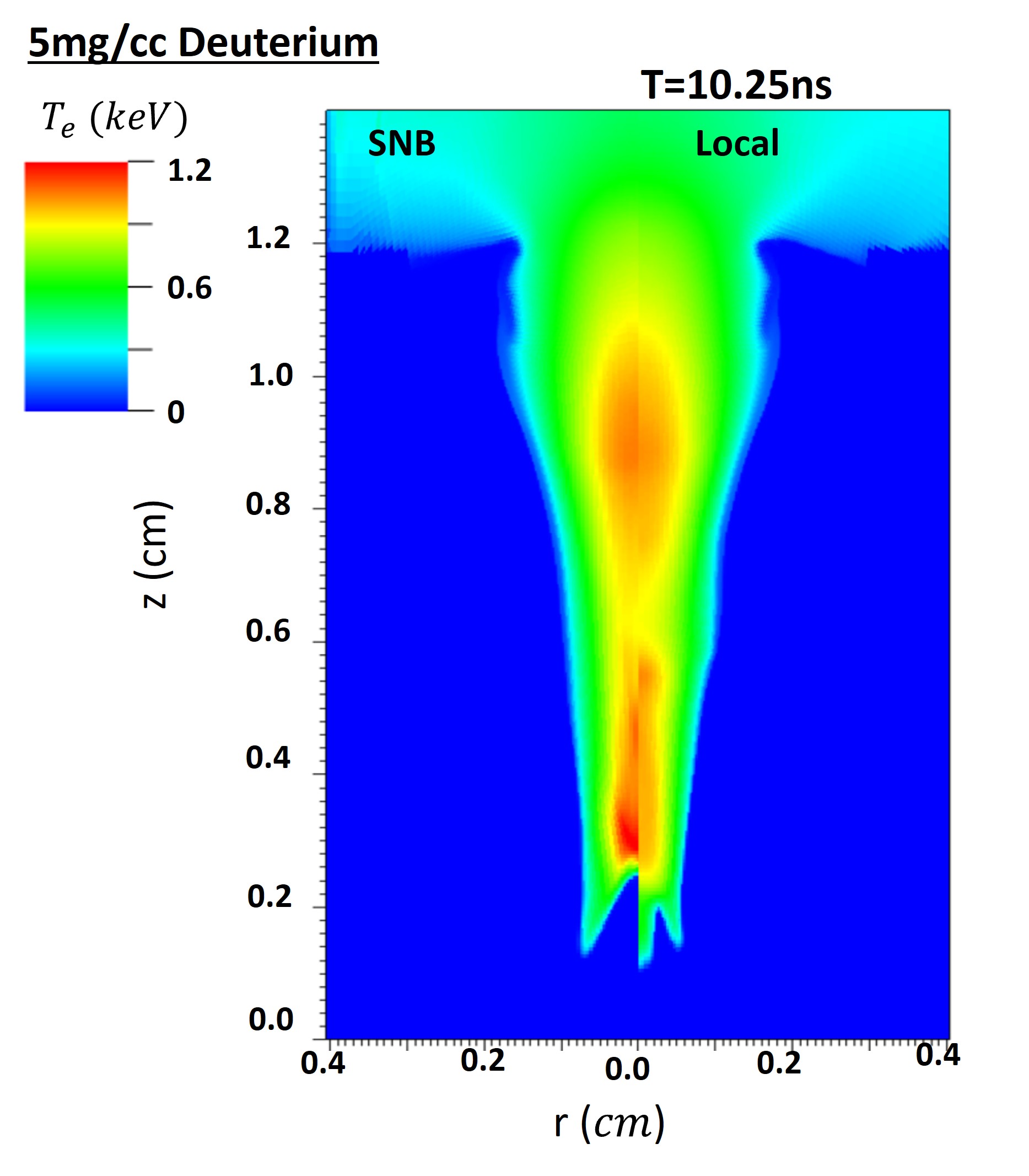} 
\caption{HYDRA simulations using SNB reduced nonlocal thermal conduction (left) and local (right) thermal conduction. 5 mg/cc of deuterium is $\approx$ 15.5\% $n_{e0} / n_{crit}$. Heat front propagation speeds do not differ significantly between the local and nonlocal evolutions.}
\label{fig: SNBD2} 

\end{figure}

Fig.\ \ref{fig: SNB_evolution} compares hydrodynamic evolution using local thermal conduction vs.\ SNB thermal conduction for neopentane. The nonlocal model increases electron temperature ($max(T_e) = 1.5$ keV) vs. the local model ($max(T_e) = 1.38$ keV). Hotter temperatures make the plasma less collisional, reducing the absorption of laser energy due to inverse bremsstrahlung. Fig. \ref{fig: SNBD2} shows HYDRA results with 5 mg/cc deuterium ($\approx$ 15.5\% $n_{e0} / n_{crit}$). Although peak temperatures in this gas profile differ by approximately 100 eV, the difference is localized primarily to the heat front and the electron temperature with either a local or nonlocal heat flow model is similar. 

Despite modest reductions in thermal conductivity shown in Fig. \ref{fig: K2C5H12}, overall temperature differences between local and nonlocal models is proportionally small. Simple estimates\cite{lindlICF} for how the electron temperature deviates due to changes in the thermal conductivity $\kappa=K_0T_e^{5/2}$ are obtained by balancing inverse bremsstrahlung absorption rates to electron thermal conduction losses at constant pressure $P_0 = nT_e$
\begin{gather}
    T_e\propto \left(\frac{1}{K_0}\right)^{1/7} \;\;.
\end{gather}
Physically nonlocal effects are reducing thermal conduction which heats up on-axis temperatures. But this reduces absorption, hence the increased laser propagation, and in turn increases conduction resulting in small overall changes to the temperature. 

Nonlocality shows a meaningful impact on $Q_{local}/Q_{K2}$ in neopentane but shows very little effect in deuterium. This is seen in figure \ref{fig: GaspipeParamSearch} where local theory over predicts the kinetic result for 3 mg/cc neopentane by a factor of 1.65x while in 5 mg/cc deuterium this is roughly 1.3x. Additionally, SNB reproduces the kinetic heat flow well for neopentane but still over predicts deuterium. This matches the local prediction more and the overall heat front propagation has little difference between a nonlocal model and a local model. 

To understand why neopentane demonstrates more nonlocality than deuterium, we return to Eq. (\ref{eq: delocLength}). Calculating $\Phi$ for both materials yields $\Phi_{\deu{}} \approx 4.195$ and $\Phi_{C_5H_{12}} \approx 1.825$ but $Z^*$ when the gas is fully ionized makes the numerical factor in the denominator roughly equal. 
In hot regions, $r \le 0.15$ cm, $n_{e,C_5H_{12}} / n_{e,\deu{}} \approx 0.75$ and $T_{e,CH} / T_{e,\deu{}} \approx 1.37$ leading to an increased $\lambda_{e,CH}^* / \lambda_{e,\deu{}}^* \approx 2.5$. The increased delocalization length in neopentane compared to deuterium is because the higher $Z^*$ material absorbs more laser energy. 
The increased laser absorption makes the neopentane gas hotter than its deuterium counterpart and $\lambda_e^* \propto T_e^2$. The kinetic heat flow between neopentane and deuterium are roughly the same order of magnitude thus the hotter neopentane gas is more nonlocal.

\section{Discussion and Conclusion}
In this paper, we examined the effects of nonlocality on thermal conduction in MagLIF gaspipes filled with neopentane and deuterium at a variety of gas fill densities. 3 mg/cc neopentane exhibited the most nonlocal effects. Using the SNB model for nonlocal electron heat conduction, which we find to replicate the radial kinetic heat flow of VFP simulations, increases the heat front propagation. This paper also explored the delocalization Knudsen number as an alternative figure of merit. The delocalization length $\lambda_e^*$ includes corrections for varying atomic number $Z$ and is a more accurate length scale for how electrons deposit their energy. Compared to a Knudsen number defined by the electron mean-free path, the delocalization Knudsen number provides a single value that represents the heat flow reduction. While Kn$_{ei} = 0.01$ gives a different heat flow reduction for neopentane and deuterium, Kn$_d$ provides a similar heat flow reduction between the different gases (see appendix). 

Plasma conditions are initialized for our K2 kinetic simulations from the radiation hydrodynamic code HYDRA. 3D effects and radiation are not included in this paper and the problem is primarily dictated by the laser propagation, hydro motion, and electron conduction. 1D radial profiles of heat flow are extracted at characteristic times and $z$ locations to capture an average effect over the full gaspipe. Full kinetic 2D simulations are desirable in the future as gradient effects are multi-dimensional, but their cost over a large enough domain is prohibitive for this study. We determined the radial heat flow to be the dominant effect after observing the axial heat flow along $z$ does not differ significantly between local, nonlocal, and kinetic heat flow models. 

Comparisons between a local heat flow model, SNB nonlocal heat flow model, and kinetic calculations from a solved electron distribution function showed that the local heat flow model consistently overpredicted SNB and kinetic models. The SNB model however, captures the nonlocal heat flow reduction well, including the preheat phenomenon. The reduction in radial heat flow leads to an increased electron temperature near $r=0$ by several hundred eV. This is only observed in neopentane which creates more nonlocal conditions. The SNB nonlocal model better captures the kinetic result in neopentane while it overpredicts the kinetic result in the deuterium case. 

Although this paper mainly investigated how electron kinetics, namely nonlocality, affects heat flow, a question for the future is how nonlocality impacts MHD effects. Magnetic fields, external or self-generated, strongly alter the electron heat flow. Sherlock and Bissell\cite{SherlockBissel2020} showed nonlocality reduces both the Biermann and Nernst terms. Walsh and Sherlock\cite{Walsh2024} provided a flux-limiter form for these terms and modifications to the electron-ion mean free path using the electron gyroradius. In both papers, it has been shown that MHD effects must be treated carefully so a unified model that accounts for nonlocal and MHD effects is needed for HED modeling.

\begin{acknowledgments}
The authors greatly appreciate helpful discussions from Chris Walsh and Joe Koning on hydrodynamic modeling and MHD effects. We thank Mehul Patel for helping simulate the gaspipes with HYDRA's SNB nonlocal package. 

This work was partly performed under the auspices of the U.S. Department of Energy by Lawrence Livermore National Laboratory under Contract No. DE-AC52-07NA27344. The information, data, or work presented herein was partly funded by Lawrence Livermore National Laboratory's Academic Collaboration Team (ACT) program.

This material is partly based upon work supported by the Department of Energy Office of Fusion Energy under Award Number(s) DE-SC0024863.

Sandia National Laboratories is a multi-mission laboratory managed and operated by National Technology \& Engineering Solutions of Sandia, LLC (NTESS), a wholly owned subsidiary of Honeywell International Inc., for the U.S. Department of Energy’s National Nuclear Security Administration (DOE/NNSA) under contract DE-NA0003525. This written work is partly authored by an employee of NTESS. The employee, not NTESS, owns the right, title and interest in and to the written work and is responsible for its contents. 

Any subjective views or opinions that might be expressed in the written work do not necessarily represent the views of the U.S. Government. The publisher acknowledges that the U.S. Government retains a non-exclusive, paid-up, irrevocable, world-wide license to publish or reproduce the published form of this written work or allow others to do so, for U.S. Government purposes. The DOE will provide public access to results of federally sponsored research in accordance with the DOE Public Access Plan.
\end{acknowledgments}

\section*{Data Availability Statement}
The data that support the findings of this study are available from the corresponding author upon reasonable request.

\appendix

\section{Figures of Merit for Nonlocal Heat Flow}
A figure of merit (FOM) for nonlocality provides a way to use general plasma profiles to predict the accuracy of local theory. Such an FOM needs to compare some collisional length scale $\lambda_\textrm{col}$ to some thermal length scale. Typically this has been the electron-ion mean free path and the temperature length scale $L_T\equiv T_e|/\nabla T_e|$. We find the delocalization length [Eq.\ \ref{eq: delocLength}], introduced by Epperlein and Short\cite{EpperleinShort}, is a more appealing FOM than the electron-ion mean free path, $\lambda_{ei}$ because it considers electron self collisions through a numerical correction. We find a nonlocal FOM where $\lambda_{\textrm{col}}=\lambda_e^*$ scales better with varying $Z$ over the FOM with $\lambda_{\textrm{col}}=\lambda_{ei}$. 

To compare the two different collisional length scales we use both the well-known "Epperlein-Short" (ES) test\cite{EpperleinShort} and a test on realistic gaspipe profiles over a finite domain. ES tests consist of small-amplitude sinusoidal temperature perturbations with a constant $n_e$ and $Z$. We initialized plasma profiles in K2 with $n_e = 5 \times 10^{20}$ cm$^{-3}$, two charge states $Z=1,13$ and spatially varying electron temperature profile $T_e = T_{e0}(1+\delta \sin kx)$ ($T_{e0}=500$ eV, $\delta=0.01$). $\lambda_{\textrm{col}}$ remains roughly constant as the plasma parameters remain constant except for the small temperature perturbation. 

After several collisional timesteps, we measure an effective thermal conductivity of the kinetic simulation $\kappa_{\textrm{K2}} \equiv |\vb{Q}|/|\nabla T_e|$ where $\vb{Q}$ is the kinetic heat flow calculated by taking moments of the evolved electron distribution function. We compare this to the Spitzer-Härm\cite{Spitzer} thermal conductivity $\kappa_{\textrm{SH}}$. A larger $k$ increases nonlocality, and reduces $\kappa_{\textrm{K2}}/\kappa_{\textrm{SH}}$. The appropriate nonlocal FOM for ES tests is $k\lambda_{\textrm{col}}$ and not the Knudsen number\cite{BellESTest} [(Eq (\ref{eq: KnudsenNumber})]. Using Kn with $L_T\equiv T_e/|\nabla T_e|$, Kn is dependent on the amplitude of the temperature perturbation, $\delta$ and space. For small $\delta$ and $\sin{kx}=1$, $L_T=\frac{1}{k\delta}\rightarrow \textrm{Kn}=\delta k \lambda_\textrm{col}$. The rate of nonlocality in this problem should only depend on $k$, for sufficiently small $\delta$.
\begin{figure}
\centering
\begin{subfigure}{0.38\textwidth} 
    \centering
    \includegraphics[width=\textwidth]{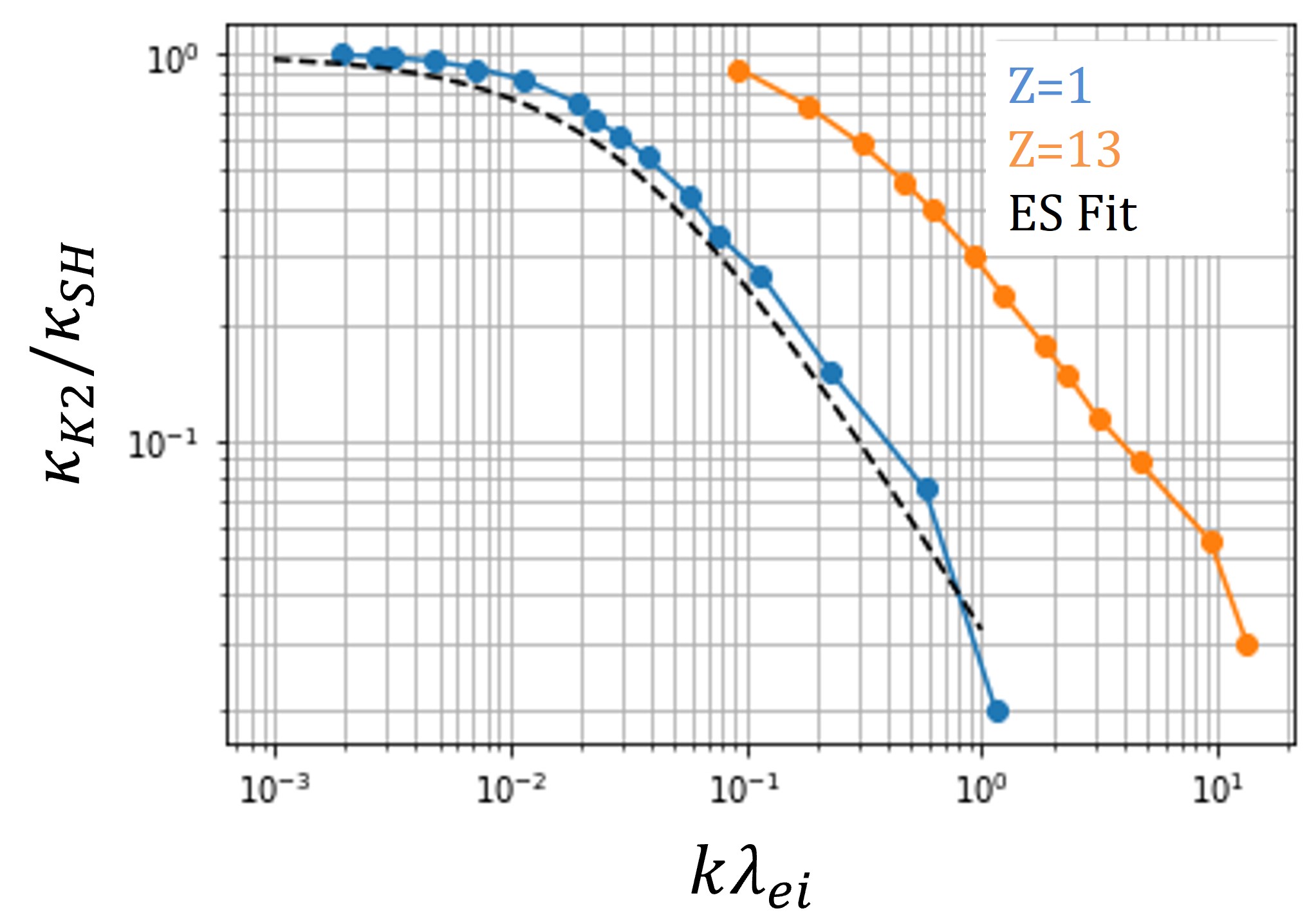}
\end{subfigure}
\hfill
\begin{subfigure}{0.47\textwidth} 
    \centering
    \includegraphics[width=\textwidth]{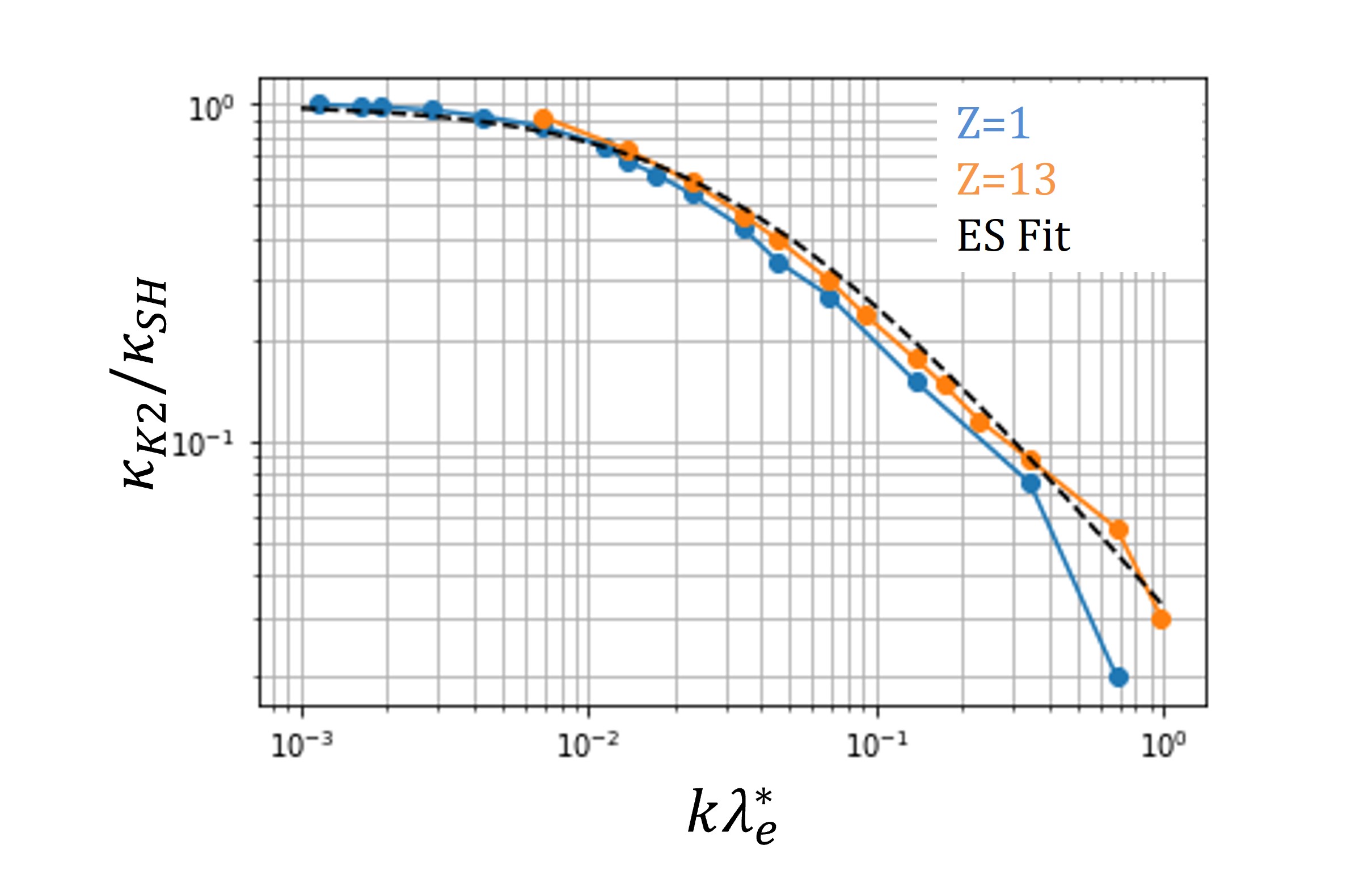}
\end{subfigure}
\hfill
\caption{Epperlein-Short tests with the K2 VFP code, for constant $n_e = 5 \times 10^{20}$ 1/cc, constant $Z=1,13$, and $T_e = T_{e0}(1+0.01 \sin kx)$ ($T_{e0}=500$ eV). The top plot's x-axis is normalized with $\lambda_{ei}$ while the bottom plot's x-axis is normalized with $\lambda_{e}^*$. Both show the reduced effective thermal conductivity($\kappa_{\textrm{K2}}$) vs. local theory($\kappa_{\textrm{SH}}$) as wavenumber increases. Solid points correspond to different ES tests with different $k$. Using $\lambda_{ei}$ produces two separated curves for varying $Z$ while using $\lambda_{e}^*$ gives a universal curve. The black dotted line is a numerical fit $\kappa_{eff} \equiv \kappa_{\textrm{K2}}/\kappa_{\textrm{SH}} = 1/(1+a(X))$ where $X$ is the respective scale length and $a=30$ fits our VFP data. }
\label{fig: EStests}
\end{figure}

Fig. \ref{fig: EStests} plots the ratio of thermal conductivities extracted from K2 and the Spitzer-Härm result from a variety of ES tests. The difference between simulations is the wavenumber $k$ of the temperature profile. Between $Z=1$ and $Z=13$, a single $k\lambda_{ei}$ cannot be used to predict the nonlocal suppressed heat flow. However, using $k\lambda_e^*$ predicts the same thermal conductivity reduction $\kappa_\textrm{eff}\equiv\kappa_{K2}/\kappa_{SH}$ and unifies the two curves. For example, $k\lambda_e^*=0.1$ predicts about a 90\% heat flow reduction for both $Z=1$ and $Z=13$. $k\lambda_{ei}=0.1$ predicts about a 80\% heat flow reduction in $Z=1$ but almost no heat flow reduction for $Z=13$. Using $\lambda_\textrm{col} = \lambda_e^*$ makes the nonlocal FOM much less sensitive to varying charge states.

\begin{figure} [h!]
\centering 
\includegraphics[scale=0.4]{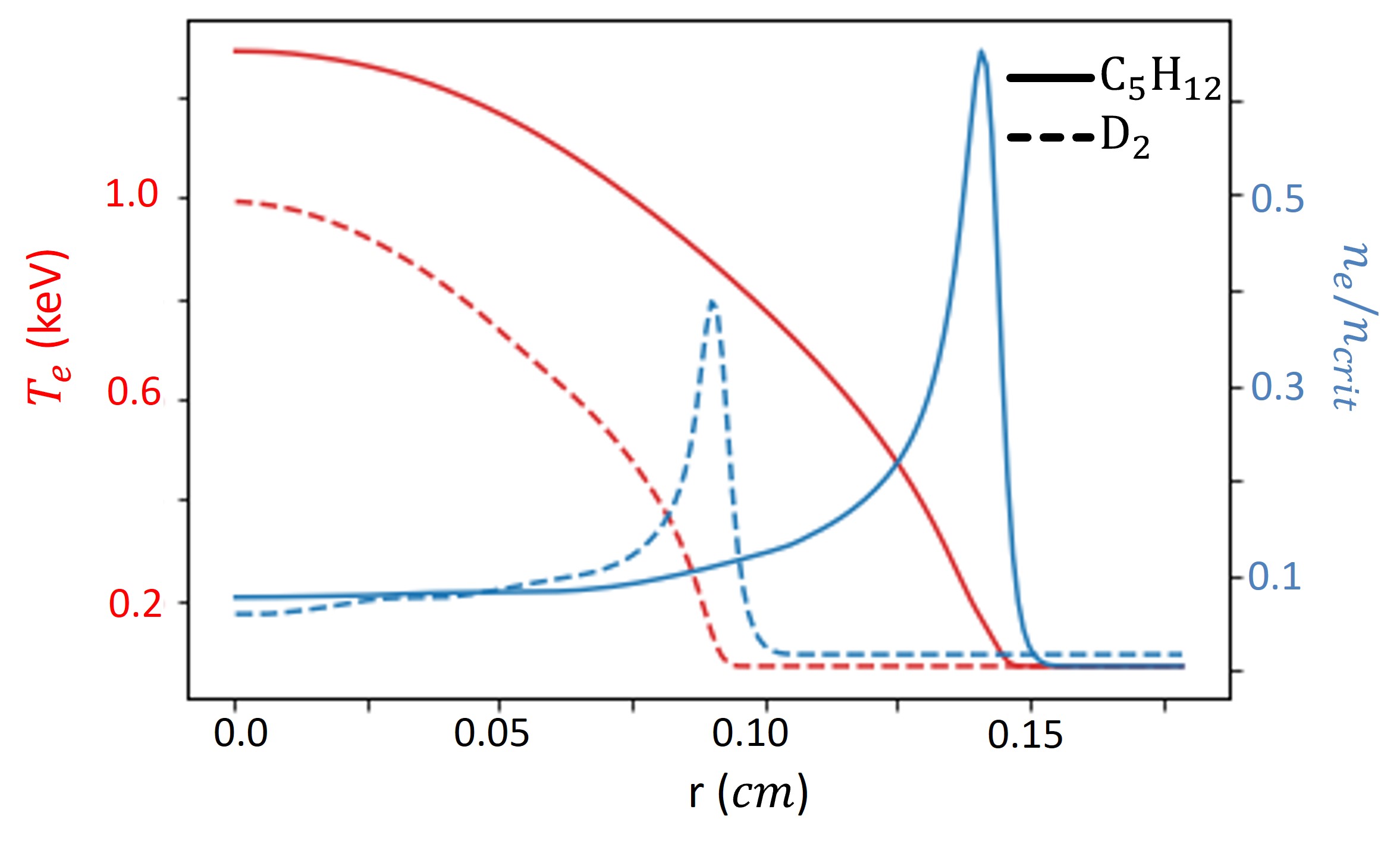} 
\caption{Gaspipe profiles used for testing nonlocal FOM Kn$\equiv \lambda_\textrm{col}/L_T$. Red shows electron temperatures while blue shows electron densities. Dashed curves correspond to \deu{} and solid curves correspond to \neop{}.}
\label{fig: NeTeES} 
\end{figure}

Our test on realistic gaspipe profiles showed a similar effect where using $\lambda_\textrm{col} = \lambda_e^*$ in a nonlocal FOM better unifies the results for \neop{} and \deu{}. For 3 mg/cc \neop{} we chose a plasma profile corresponding to $t=8$ ns at $z=0.65$ cm and for 5mg/cc \deu{} we chose a plasma profile corresponding to $t=10$ ns at $z=0.6$ cm, as shown in Fig. \ref{fig: NeTeES}. These profiles had a smooth $\vb{Q}_\perp$ making them ideal for our test cases. 

For these gaspipe tests we keep $n_e$ and $Z$ profiles constant and multiply $T_e$ by a numerical factor to increase thermal gradients and thus increase nonlocal effects. The appropriate nonlocal FOM becomes the Knudsen number $\equiv \lambda_\textrm{col}/L_T$ for non-periodic profiles. 

\begin{figure}
\centering
\begin{subfigure}{0.40\textwidth} 
    \centering
    \includegraphics[width=\textwidth]{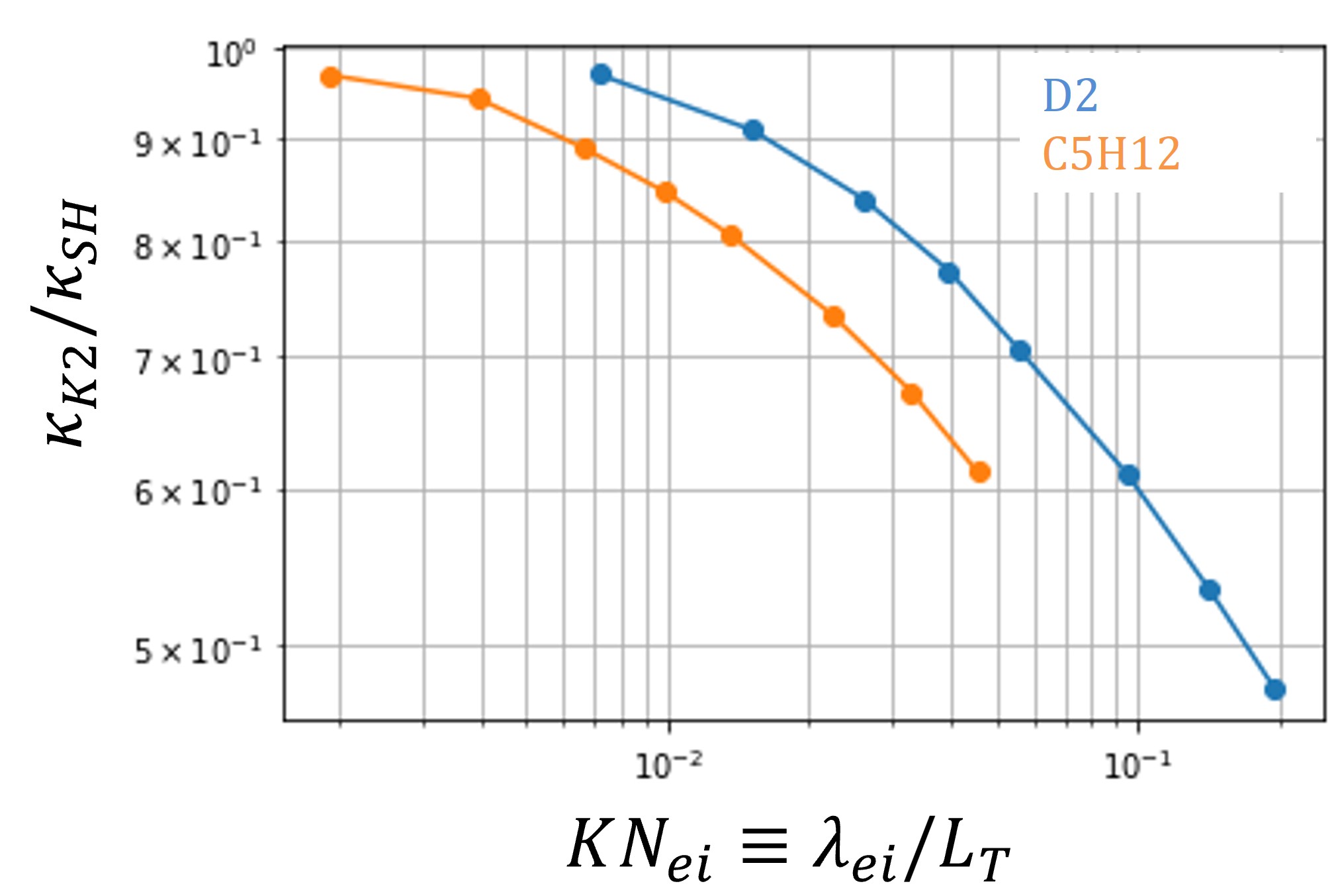}
\end{subfigure}
\hfill
\begin{subfigure}{0.45\textwidth} 
    \centering
    \includegraphics[width=\textwidth]{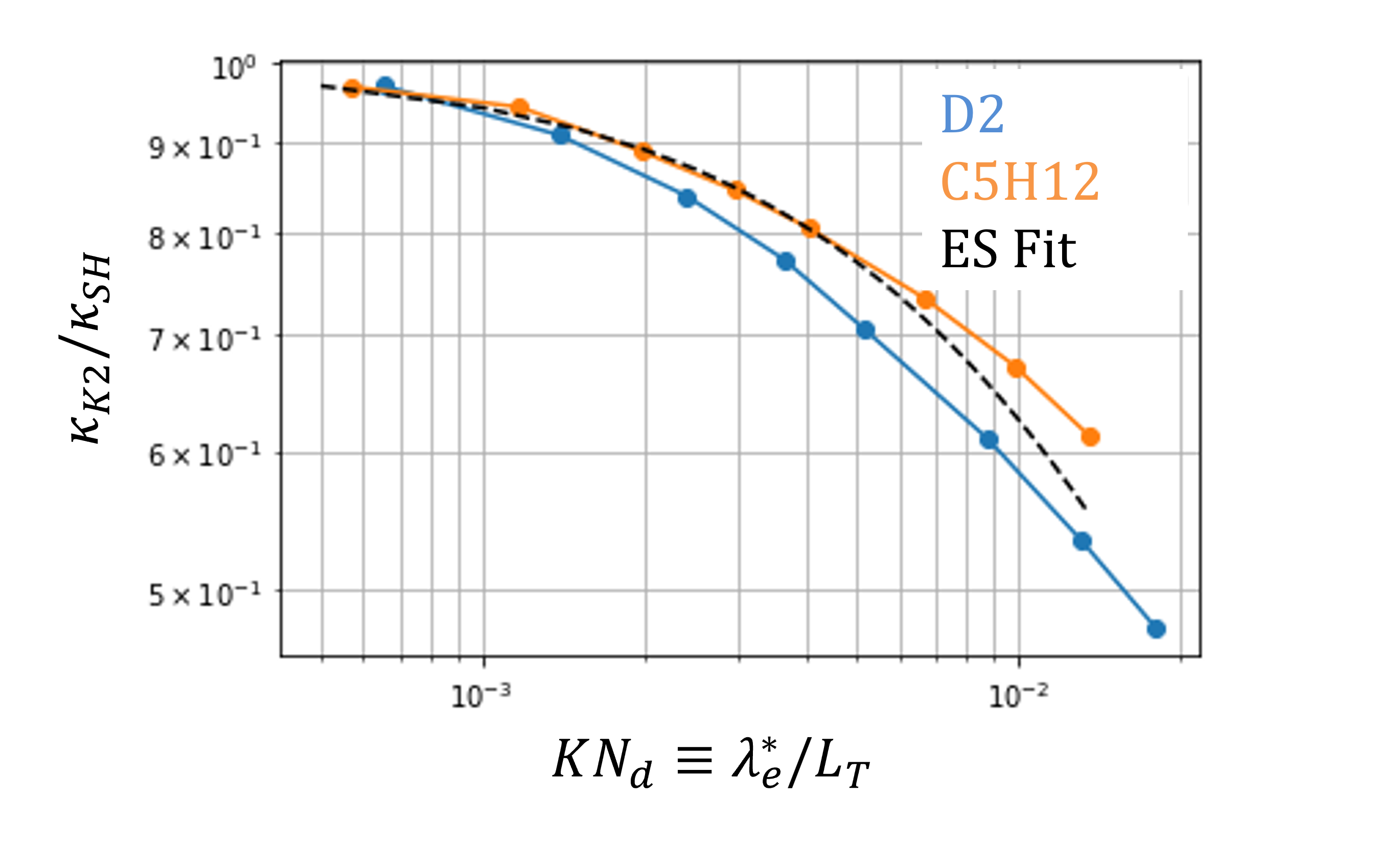}
\end{subfigure}
\hfill
\caption{Epperlein-Short tests with K2 on realistic gaspipe profile tests for Kn are shown. $T_e$ was multiplied by a numerical factor to induce stronger nonlocal effects. Using a nonlocal FOM where $\lambda_\textrm{col}=\lambda_e^*$ (Bottom) predicts a similar heat flow reduction for varying gas fills (\deu{} and \neop{}) as opposed to $\lambda_\textrm{col}=\lambda_{ei}$ (Top). A numerical fit $\kappa_{eff} = \kappa_{K2}/\kappa_{SH}=1/(1+a(\lambda_e^*/L_T))$ where $a=60$ fits our VFP data well.}
\label{fig: Gaspipetests}
\end{figure}

Fig. \ref{fig: Gaspipetests} shows that Kn$_d$($\lambda_\textrm{col}=\lambda_e^*$) better unifies the results between \deu{} and \neop{} than using Kn$_{ei}$($\lambda_\textrm{col}=\lambda_{ei}$). Similar to the ES tests, using the delocalization length in our nonlocal FOM provides more consistent predictions than using the electron-ion mean free path. A numerical fit $\kappa_{eff} = 1/(1+a(\lambda_e^*/L_T))$ where $a=60$ allows us to probe gaspipe data for where nonlocality might have impacts on the overall dynamics. In other words, regions where Kn$_d \ge 0.3\% \rightarrow \kappa_{eff}\lesssim 84.7\%$ should be investigated using K2 for potentially large kinetic effects.

\nocite{*}
\bibliography{gaspipe}

\providecommand{\noopsort}[1]{}\providecommand{\singleletter}[1]{#1}%
\begin{thebibliography}{46}%
\makeatletter
\providecommand \@ifxundefined [1]{%
 \@ifx{#1\undefined}
}%
\providecommand \@ifnum [1]{%
 \ifnum #1\expandafter \@firstoftwo
 \else \expandafter \@secondoftwo
 \fi
}%
\providecommand \@ifx [1]{%
 \ifx #1\expandafter \@firstoftwo
 \else \expandafter \@secondoftwo
 \fi
}%
\providecommand \natexlab [1]{#1}%
\providecommand \enquote  [1]{``#1''}%
\providecommand \bibnamefont  [1]{#1}%
\providecommand \bibfnamefont [1]{#1}%
\providecommand \citenamefont [1]{#1}%
\providecommand \href@noop [0]{\@secondoftwo}%
\providecommand \href [0]{\begingroup \@sanitize@url \@href}%
\providecommand \@href[1]{\@@startlink{#1}\@@href}%
\providecommand \@@href[1]{\endgroup#1\@@endlink}%
\providecommand \@sanitize@url [0]{\catcode `\\12\catcode `\$12\catcode `\&12\catcode `\#12\catcode `\^12\catcode `\_12\catcode `\%12\relax}%
\providecommand \@@startlink[1]{}%
\providecommand \@@endlink[0]{}%
\providecommand \url  [0]{\begingroup\@sanitize@url \@url }%
\providecommand \@url [1]{\endgroup\@href {#1}{\urlprefix }}%
\providecommand \urlprefix  [0]{URL }%
\providecommand \Eprint [0]{\href }%
\providecommand \doibase [0]{http://dx.doi.org/}%
\providecommand \selectlanguage [0]{\@gobble}%
\providecommand \bibinfo  [0]{\@secondoftwo}%
\providecommand \bibfield  [0]{\@secondoftwo}%
\providecommand \translation [1]{[#1]}%
\providecommand \BibitemOpen [0]{}%
\providecommand \bibitemStop [0]{}%
\providecommand \bibitemNoStop [0]{.\EOS\space}%
\providecommand \EOS [0]{\spacefactor3000\relax}%
\providecommand \BibitemShut  [1]{\csname bibitem#1\endcsname}%
\let\auto@bib@innerbib\@empty
\bibitem [{\citenamefont {Moses}\ \emph {et~al.}(2016)\citenamefont {Moses}, \citenamefont {Lindl}, \citenamefont {Spaeth}, \citenamefont {Patterson}, \citenamefont {Sawicki}, \citenamefont {Atherton}, \citenamefont {Baisden}, \citenamefont {Lagin}, \citenamefont {Larson}, \citenamefont {MacGowan}, \citenamefont {Miller}, \citenamefont {Rardin}, \citenamefont {Roberts}, \citenamefont {Wonterghem},\ and\ \citenamefont {Wegner}}]{NIFref}%
  \BibitemOpen
  \bibfield  {author} {\bibinfo {author} {\bibfnamefont {E.~I.}\ \bibnamefont {Moses}}, \bibinfo {author} {\bibfnamefont {J.~D.}\ \bibnamefont {Lindl}}, \bibinfo {author} {\bibfnamefont {M.~L.}\ \bibnamefont {Spaeth}}, \bibinfo {author} {\bibfnamefont {R.~W.}\ \bibnamefont {Patterson}}, \bibinfo {author} {\bibfnamefont {R.~H.}\ \bibnamefont {Sawicki}}, \bibinfo {author} {\bibfnamefont {L.~J.}\ \bibnamefont {Atherton}}, \bibinfo {author} {\bibfnamefont {P.~A.}\ \bibnamefont {Baisden}}, \bibinfo {author} {\bibfnamefont {L.~J.}\ \bibnamefont {Lagin}}, \bibinfo {author} {\bibfnamefont {D.~W.}\ \bibnamefont {Larson}}, \bibinfo {author} {\bibfnamefont {B.~J.}\ \bibnamefont {MacGowan}}, \bibinfo {author} {\bibfnamefont {G.~H.}\ \bibnamefont {Miller}}, \bibinfo {author} {\bibfnamefont {D.~C.}\ \bibnamefont {Rardin}}, \bibinfo {author} {\bibfnamefont {V.~S.}\ \bibnamefont {Roberts}}, \bibinfo {author} {\bibfnamefont {B.~M.~V.}\ \bibnamefont {Wonterghem}}, \ and\ \bibinfo {author} {\bibfnamefont {P.~J.}\ \bibnamefont
  {Wegner}},\ }\bibfield  {title} {\enquote {\bibinfo {title} {{Overview: Development of the National Ignition Facility and the Transition to a User Facility for the Ignition Campaign and High Energy Density Scientific Research}},}\ }\href {\doibase 10.13182/FST15-128} {\bibfield  {journal} {\bibinfo  {journal} {Fusion Science and Technology}\ }\textbf {\bibinfo {volume} {69}},\ \bibinfo {pages} {1--24} (\bibinfo {year} {2016})},\ \Eprint {http://arxiv.org/abs/https://doi.org/10.13182/FST15-128} {https://doi.org/10.13182/FST15-128} \BibitemShut {NoStop}%
\bibitem [{\citenamefont {Slutz}\ and\ \citenamefont {Vesey}(2012)}]{Slutz2012}%
  \BibitemOpen
  \bibfield  {author} {\bibinfo {author} {\bibfnamefont {S.~A.}\ \bibnamefont {Slutz}}\ and\ \bibinfo {author} {\bibfnamefont {R.~A.}\ \bibnamefont {Vesey}},\ }\bibfield  {title} {\enquote {\bibinfo {title} {High-gain magnetized inertial fusion},}\ }\href {\doibase 10.1103/PhysRevLett.108.025003} {\bibfield  {journal} {\bibinfo  {journal} {Phys. Rev. Lett.}\ }\textbf {\bibinfo {volume} {108}},\ \bibinfo {pages} {025003} (\bibinfo {year} {2012})}\BibitemShut {NoStop}%
\bibitem [{\citenamefont {Nuckolls}\ \emph {et~al.}(1972)\citenamefont {Nuckolls}, \citenamefont {Wood}, \citenamefont {Thiessen},\ and\ \citenamefont {Zimmerman}}]{directdrive}%
  \BibitemOpen
  \bibfield  {author} {\bibinfo {author} {\bibfnamefont {J.~H.}\ \bibnamefont {Nuckolls}}, \bibinfo {author} {\bibfnamefont {L.}~\bibnamefont {Wood}}, \bibinfo {author} {\bibfnamefont {A.}~\bibnamefont {Thiessen}}, \ and\ \bibinfo {author} {\bibfnamefont {G.~B.}\ \bibnamefont {Zimmerman}},\ }\bibfield  {title} {\enquote {\bibinfo {title} {Laser compression of matter to super-high densities: Thermonuclear (ctr) applications},}\ }\href@noop {} {\bibfield  {journal} {\bibinfo  {journal} {Nature}\ }\textbf {\bibinfo {volume} {239}},\ \bibinfo {pages} {139--142} (\bibinfo {year} {1972})}\BibitemShut {NoStop}%
\bibitem [{\citenamefont {Lindl}(1995{\natexlab{a}})}]{indirectdrive}%
  \BibitemOpen
  \bibfield  {author} {\bibinfo {author} {\bibfnamefont {J.}~\bibnamefont {Lindl}},\ }\bibfield  {title} {\enquote {\bibinfo {title} {{Development of the indirect‐drive approach to inertial confinement fusion and the target physics basis for ignition and gain}},}\ }\href {\doibase 10.1063/1.871025} {\bibfield  {journal} {\bibinfo  {journal} {Physics of Plasmas}\ }\textbf {\bibinfo {volume} {2}},\ \bibinfo {pages} {3933--4024} (\bibinfo {year} {1995}{\natexlab{a}})},\ \Eprint {http://arxiv.org/abs/https://pubs.aip.org/aip/pop/article-pdf/2/11/3933/19277171/3933\_1\_online.pdf} {https://pubs.aip.org/aip/pop/article-pdf/2/11/3933/19277171/3933\_1\_online.pdf} \BibitemShut {NoStop}%
\bibitem [{\citenamefont {Slutz}\ \emph {et~al.}(2018)\citenamefont {Slutz}, \citenamefont {Gomez}, \citenamefont {Hansen}, \citenamefont {Harding}, \citenamefont {Hutsel}, \citenamefont {Knapp}, \citenamefont {Lamppa}, \citenamefont {Awe}, \citenamefont {Ampleford}, \citenamefont {Bliss}, \citenamefont {Chandler}, \citenamefont {Cuneo}, \citenamefont {Geissel}, \citenamefont {Glinsky}, \citenamefont {Harvey-Thompson}, \citenamefont {Hess}, \citenamefont {Jennings}, \citenamefont {Jones}, \citenamefont {Laity}, \citenamefont {Martin}, \citenamefont {Peterson}, \citenamefont {Porter}, \citenamefont {Rambo}, \citenamefont {Rochau}, \citenamefont {Ruiz}, \citenamefont {Savage}, \citenamefont {Schwarz}, \citenamefont {Schmit}, \citenamefont {Shipley}, \citenamefont {Sinars}, \citenamefont {Smith}, \citenamefont {Vesey},\ and\ \citenamefont {Weis}}]{Slutz2018}%
  \BibitemOpen
  \bibfield  {author} {\bibinfo {author} {\bibfnamefont {S.~A.}\ \bibnamefont {Slutz}}, \bibinfo {author} {\bibfnamefont {M.~R.}\ \bibnamefont {Gomez}}, \bibinfo {author} {\bibfnamefont {S.~B.}\ \bibnamefont {Hansen}}, \bibinfo {author} {\bibfnamefont {E.~C.}\ \bibnamefont {Harding}}, \bibinfo {author} {\bibfnamefont {B.~T.}\ \bibnamefont {Hutsel}}, \bibinfo {author} {\bibfnamefont {P.~F.}\ \bibnamefont {Knapp}}, \bibinfo {author} {\bibfnamefont {D.~C.}\ \bibnamefont {Lamppa}}, \bibinfo {author} {\bibfnamefont {T.~J.}\ \bibnamefont {Awe}}, \bibinfo {author} {\bibfnamefont {D.~J.}\ \bibnamefont {Ampleford}}, \bibinfo {author} {\bibfnamefont {D.~E.}\ \bibnamefont {Bliss}}, \bibinfo {author} {\bibfnamefont {G.~A.}\ \bibnamefont {Chandler}}, \bibinfo {author} {\bibfnamefont {M.~E.}\ \bibnamefont {Cuneo}}, \bibinfo {author} {\bibfnamefont {M.}~\bibnamefont {Geissel}}, \bibinfo {author} {\bibfnamefont {M.~E.}\ \bibnamefont {Glinsky}}, \bibinfo {author} {\bibfnamefont {A.~J.}\ \bibnamefont {Harvey-Thompson}},
  \bibinfo {author} {\bibfnamefont {M.~H.}\ \bibnamefont {Hess}}, \bibinfo {author} {\bibfnamefont {C.~A.}\ \bibnamefont {Jennings}}, \bibinfo {author} {\bibfnamefont {B.}~\bibnamefont {Jones}}, \bibinfo {author} {\bibfnamefont {G.~R.}\ \bibnamefont {Laity}}, \bibinfo {author} {\bibfnamefont {M.~R.}\ \bibnamefont {Martin}}, \bibinfo {author} {\bibfnamefont {K.~J.}\ \bibnamefont {Peterson}}, \bibinfo {author} {\bibfnamefont {J.~L.}\ \bibnamefont {Porter}}, \bibinfo {author} {\bibfnamefont {P.~K.}\ \bibnamefont {Rambo}}, \bibinfo {author} {\bibfnamefont {G.~A.}\ \bibnamefont {Rochau}}, \bibinfo {author} {\bibfnamefont {C.~L.}\ \bibnamefont {Ruiz}}, \bibinfo {author} {\bibfnamefont {M.~E.}\ \bibnamefont {Savage}}, \bibinfo {author} {\bibfnamefont {J.}~\bibnamefont {Schwarz}}, \bibinfo {author} {\bibfnamefont {P.~F.}\ \bibnamefont {Schmit}}, \bibinfo {author} {\bibfnamefont {G.}~\bibnamefont {Shipley}}, \bibinfo {author} {\bibfnamefont {D.~B.}\ \bibnamefont {Sinars}}, \bibinfo {author} {\bibfnamefont {I.~C.}\
  \bibnamefont {Smith}}, \bibinfo {author} {\bibfnamefont {R.~A.}\ \bibnamefont {Vesey}}, \ and\ \bibinfo {author} {\bibfnamefont {M.~R.}\ \bibnamefont {Weis}},\ }\bibfield  {title} {\enquote {\bibinfo {title} {{Enhancing performance of magnetized liner inertial fusion at the Z facility}},}\ }\href {\doibase 10.1063/1.5054317} {\bibfield  {journal} {\bibinfo  {journal} {Physics of Plasmas}\ }\textbf {\bibinfo {volume} {25}},\ \bibinfo {pages} {112706} (\bibinfo {year} {2018})},\ \Eprint {http://arxiv.org/abs/https://pubs.aip.org/aip/pop/article-pdf/doi/10.1063/1.5054317/15954352/112706\_1\_online.pdf} {https://pubs.aip.org/aip/pop/article-pdf/doi/10.1063/1.5054317/15954352/112706\_1\_online.pdf} \BibitemShut {NoStop}%
\bibitem [{\citenamefont {Sefkow}\ \emph {et~al.}(2014)\citenamefont {Sefkow}, \citenamefont {Slutz}, \citenamefont {Koning}, \citenamefont {Marinak}, \citenamefont {Peterson}, \citenamefont {Sinars},\ and\ \citenamefont {Vesey}}]{Sefkow2014}%
  \BibitemOpen
  \bibfield  {author} {\bibinfo {author} {\bibfnamefont {A.~B.}\ \bibnamefont {Sefkow}}, \bibinfo {author} {\bibfnamefont {S.~A.}\ \bibnamefont {Slutz}}, \bibinfo {author} {\bibfnamefont {J.~M.}\ \bibnamefont {Koning}}, \bibinfo {author} {\bibfnamefont {M.~M.}\ \bibnamefont {Marinak}}, \bibinfo {author} {\bibfnamefont {K.~J.}\ \bibnamefont {Peterson}}, \bibinfo {author} {\bibfnamefont {D.~B.}\ \bibnamefont {Sinars}}, \ and\ \bibinfo {author} {\bibfnamefont {R.~A.}\ \bibnamefont {Vesey}},\ }\bibfield  {title} {\enquote {\bibinfo {title} {{Design of magnetized liner inertial fusion experiments using the Z facility}},}\ }\href {\doibase 10.1063/1.4890298} {\bibfield  {journal} {\bibinfo  {journal} {Physics of Plasmas}\ }\textbf {\bibinfo {volume} {21}},\ \bibinfo {pages} {072711} (\bibinfo {year} {2014})},\ \Eprint {http://arxiv.org/abs/https://pubs.aip.org/aip/pop/article-pdf/doi/10.1063/1.4890298/15694058/072711\_1\_online.pdf}
  {https://pubs.aip.org/aip/pop/article-pdf/doi/10.1063/1.4890298/15694058/072711\_1\_online.pdf} \BibitemShut {NoStop}%
\bibitem [{\citenamefont {Harvey-Thompson}\ \emph {et~al.}(2019)\citenamefont {Harvey-Thompson}, \citenamefont {Geissel}, \citenamefont {Jennings}, \citenamefont {Weis}, \citenamefont {Gomez}, \citenamefont {Fein}, \citenamefont {Ampleford}, \citenamefont {Chandler}, \citenamefont {Glinsky}, \citenamefont {Hahn}, \citenamefont {Hansen}, \citenamefont {Harding}, \citenamefont {Knapp}, \citenamefont {Paguio}, \citenamefont {Perea}, \citenamefont {Peterson}, \citenamefont {Porter}, \citenamefont {Rambo}, \citenamefont {Robertson}, \citenamefont {Rochau}, \citenamefont {Ruiz}, \citenamefont {Schwarz}, \citenamefont {Shores}, \citenamefont {Sinars}, \citenamefont {Slutz}, \citenamefont {Smith}, \citenamefont {Smith}, \citenamefont {Speas}, \citenamefont {Whittemore},\ and\ \citenamefont {Woodbury}}]{Harvey-Thompson2019}%
  \BibitemOpen
  \bibfield  {author} {\bibinfo {author} {\bibfnamefont {A.~J.}\ \bibnamefont {Harvey-Thompson}}, \bibinfo {author} {\bibfnamefont {M.}~\bibnamefont {Geissel}}, \bibinfo {author} {\bibfnamefont {C.~A.}\ \bibnamefont {Jennings}}, \bibinfo {author} {\bibfnamefont {M.~R.}\ \bibnamefont {Weis}}, \bibinfo {author} {\bibfnamefont {M.~R.}\ \bibnamefont {Gomez}}, \bibinfo {author} {\bibfnamefont {J.~R.}\ \bibnamefont {Fein}}, \bibinfo {author} {\bibfnamefont {D.~J.}\ \bibnamefont {Ampleford}}, \bibinfo {author} {\bibfnamefont {G.~A.}\ \bibnamefont {Chandler}}, \bibinfo {author} {\bibfnamefont {M.~E.}\ \bibnamefont {Glinsky}}, \bibinfo {author} {\bibfnamefont {K.~D.}\ \bibnamefont {Hahn}}, \bibinfo {author} {\bibfnamefont {S.~B.}\ \bibnamefont {Hansen}}, \bibinfo {author} {\bibfnamefont {E.~C.}\ \bibnamefont {Harding}}, \bibinfo {author} {\bibfnamefont {P.~F.}\ \bibnamefont {Knapp}}, \bibinfo {author} {\bibfnamefont {R.~R.}\ \bibnamefont {Paguio}}, \bibinfo {author} {\bibfnamefont {L.}~\bibnamefont {Perea}}, \bibinfo
  {author} {\bibfnamefont {K.~J.}\ \bibnamefont {Peterson}}, \bibinfo {author} {\bibfnamefont {J.~L.}\ \bibnamefont {Porter}}, \bibinfo {author} {\bibfnamefont {P.~K.}\ \bibnamefont {Rambo}}, \bibinfo {author} {\bibfnamefont {G.~K.}\ \bibnamefont {Robertson}}, \bibinfo {author} {\bibfnamefont {G.~A.}\ \bibnamefont {Rochau}}, \bibinfo {author} {\bibfnamefont {C.~L.}\ \bibnamefont {Ruiz}}, \bibinfo {author} {\bibfnamefont {J.}~\bibnamefont {Schwarz}}, \bibinfo {author} {\bibfnamefont {J.~E.}\ \bibnamefont {Shores}}, \bibinfo {author} {\bibfnamefont {D.~B.}\ \bibnamefont {Sinars}}, \bibinfo {author} {\bibfnamefont {S.~A.}\ \bibnamefont {Slutz}}, \bibinfo {author} {\bibfnamefont {G.~E.}\ \bibnamefont {Smith}}, \bibinfo {author} {\bibfnamefont {I.~C.}\ \bibnamefont {Smith}}, \bibinfo {author} {\bibfnamefont {C.~S.}\ \bibnamefont {Speas}}, \bibinfo {author} {\bibfnamefont {K.}~\bibnamefont {Whittemore}}, \ and\ \bibinfo {author} {\bibfnamefont {D.}~\bibnamefont {Woodbury}},\ }\bibfield  {title} {\enquote {\bibinfo
  {title} {{Constraining preheat energy deposition in MagLIF experiments with multi-frame shadowgraphy}},}\ }\href {\doibase 10.1063/1.5086044} {\bibfield  {journal} {\bibinfo  {journal} {Physics of Plasmas}\ }\textbf {\bibinfo {volume} {26}},\ \bibinfo {pages} {032707} (\bibinfo {year} {2019})},\ \Eprint {http://arxiv.org/abs/https://pubs.aip.org/aip/pop/article-pdf/doi/10.1063/1.5086044/15833685/032707\_1\_online.pdf} {https://pubs.aip.org/aip/pop/article-pdf/doi/10.1063/1.5086044/15833685/032707\_1\_online.pdf} \BibitemShut {NoStop}%
\bibitem [{\citenamefont {Gomez}\ \emph {et~al.}(2014)\citenamefont {Gomez}, \citenamefont {Slutz}, \citenamefont {Sefkow}, \citenamefont {Sinars}, \citenamefont {Hahn}, \citenamefont {Hansen}, \citenamefont {Harding}, \citenamefont {Knapp}, \citenamefont {Schmit}, \citenamefont {Jennings}, \citenamefont {Awe}, \citenamefont {Geissel}, \citenamefont {Rovang}, \citenamefont {Chandler}, \citenamefont {Cooper}, \citenamefont {Cuneo}, \citenamefont {Harvey-Thompson}, \citenamefont {Herrmann}, \citenamefont {Hess}, \citenamefont {Johns}, \citenamefont {Lamppa}, \citenamefont {Martin}, \citenamefont {McBride}, \citenamefont {Peterson}, \citenamefont {Porter}, \citenamefont {Robertson}, \citenamefont {Rochau}, \citenamefont {Ruiz}, \citenamefont {Savage}, \citenamefont {Smith}, \citenamefont {Stygar},\ and\ \citenamefont {Vesey}}]{Gomez2014}%
  \BibitemOpen
  \bibfield  {author} {\bibinfo {author} {\bibfnamefont {M.~R.}\ \bibnamefont {Gomez}}, \bibinfo {author} {\bibfnamefont {S.~A.}\ \bibnamefont {Slutz}}, \bibinfo {author} {\bibfnamefont {A.~B.}\ \bibnamefont {Sefkow}}, \bibinfo {author} {\bibfnamefont {D.~B.}\ \bibnamefont {Sinars}}, \bibinfo {author} {\bibfnamefont {K.~D.}\ \bibnamefont {Hahn}}, \bibinfo {author} {\bibfnamefont {S.~B.}\ \bibnamefont {Hansen}}, \bibinfo {author} {\bibfnamefont {E.~C.}\ \bibnamefont {Harding}}, \bibinfo {author} {\bibfnamefont {P.~F.}\ \bibnamefont {Knapp}}, \bibinfo {author} {\bibfnamefont {P.~F.}\ \bibnamefont {Schmit}}, \bibinfo {author} {\bibfnamefont {C.~A.}\ \bibnamefont {Jennings}}, \bibinfo {author} {\bibfnamefont {T.~J.}\ \bibnamefont {Awe}}, \bibinfo {author} {\bibfnamefont {M.}~\bibnamefont {Geissel}}, \bibinfo {author} {\bibfnamefont {D.~C.}\ \bibnamefont {Rovang}}, \bibinfo {author} {\bibfnamefont {G.~A.}\ \bibnamefont {Chandler}}, \bibinfo {author} {\bibfnamefont {G.~W.}\ \bibnamefont {Cooper}}, \bibinfo {author}
  {\bibfnamefont {M.~E.}\ \bibnamefont {Cuneo}}, \bibinfo {author} {\bibfnamefont {A.~J.}\ \bibnamefont {Harvey-Thompson}}, \bibinfo {author} {\bibfnamefont {M.~C.}\ \bibnamefont {Herrmann}}, \bibinfo {author} {\bibfnamefont {M.~H.}\ \bibnamefont {Hess}}, \bibinfo {author} {\bibfnamefont {O.}~\bibnamefont {Johns}}, \bibinfo {author} {\bibfnamefont {D.~C.}\ \bibnamefont {Lamppa}}, \bibinfo {author} {\bibfnamefont {M.~R.}\ \bibnamefont {Martin}}, \bibinfo {author} {\bibfnamefont {R.~D.}\ \bibnamefont {McBride}}, \bibinfo {author} {\bibfnamefont {K.~J.}\ \bibnamefont {Peterson}}, \bibinfo {author} {\bibfnamefont {J.~L.}\ \bibnamefont {Porter}}, \bibinfo {author} {\bibfnamefont {G.~K.}\ \bibnamefont {Robertson}}, \bibinfo {author} {\bibfnamefont {G.~A.}\ \bibnamefont {Rochau}}, \bibinfo {author} {\bibfnamefont {C.~L.}\ \bibnamefont {Ruiz}}, \bibinfo {author} {\bibfnamefont {M.~E.}\ \bibnamefont {Savage}}, \bibinfo {author} {\bibfnamefont {I.~C.}\ \bibnamefont {Smith}}, \bibinfo {author} {\bibfnamefont {W.~A.}\
  \bibnamefont {Stygar}}, \ and\ \bibinfo {author} {\bibfnamefont {R.~A.}\ \bibnamefont {Vesey}},\ }\bibfield  {title} {\enquote {\bibinfo {title} {Experimental demonstration of fusion-relevant conditions in magnetized liner inertial fusion},}\ }\href {\doibase 10.1103/PhysRevLett.113.155003} {\bibfield  {journal} {\bibinfo  {journal} {Phys. Rev. Lett.}\ }\textbf {\bibinfo {volume} {113}},\ \bibinfo {pages} {155003} (\bibinfo {year} {2014})}\BibitemShut {NoStop}%
\bibitem [{\citenamefont {Cuneo}\ \emph {et~al.}(2012)\citenamefont {Cuneo}, \citenamefont {Herrmann}, \citenamefont {Sinars}, \citenamefont {Slutz}, \citenamefont {Stygar}, \citenamefont {Vesey}, \citenamefont {Sefkow}, \citenamefont {Rochau}, \citenamefont {Chandler}, \citenamefont {Bailey}, \citenamefont {Porter}, \citenamefont {McBride}, \citenamefont {Rovang}, \citenamefont {Mazarakis}, \citenamefont {Yu}, \citenamefont {Lamppa}, \citenamefont {Peterson}, \citenamefont {Nakhleh}, \citenamefont {Hansen}, \citenamefont {Lopez}, \citenamefont {Savage}, \citenamefont {Jennings}, \citenamefont {Martin}, \citenamefont {Lemke}, \citenamefont {Atherton}, \citenamefont {Smith}, \citenamefont {Rambo}, \citenamefont {Jones}, \citenamefont {Lopez}, \citenamefont {Christenson}, \citenamefont {Sweeney}, \citenamefont {Jones}, \citenamefont {McPherson}, \citenamefont {Harding}, \citenamefont {Gomez}, \citenamefont {Knapp}, \citenamefont {Awe}, \citenamefont {Leeper}, \citenamefont {Ruiz}, \citenamefont {Cooper},
  \citenamefont {Hahn}, \citenamefont {McKenney}, \citenamefont {Owen}, \citenamefont {McKee}, \citenamefont {Leifeste}, \citenamefont {Ampleford}, \citenamefont {Waisman}, \citenamefont {Harvey-Thompson}, \citenamefont {Kaye}, \citenamefont {Hess}, \citenamefont {Rosenthal},\ and\ \citenamefont {Matzen}}]{Cuneo2012}%
  \BibitemOpen
  \bibfield  {author} {\bibinfo {author} {\bibfnamefont {M.~E.}\ \bibnamefont {Cuneo}}, \bibinfo {author} {\bibfnamefont {M.~C.}\ \bibnamefont {Herrmann}}, \bibinfo {author} {\bibfnamefont {D.~B.}\ \bibnamefont {Sinars}}, \bibinfo {author} {\bibfnamefont {S.~A.}\ \bibnamefont {Slutz}}, \bibinfo {author} {\bibfnamefont {W.~A.}\ \bibnamefont {Stygar}}, \bibinfo {author} {\bibfnamefont {R.~A.}\ \bibnamefont {Vesey}}, \bibinfo {author} {\bibfnamefont {A.~B.}\ \bibnamefont {Sefkow}}, \bibinfo {author} {\bibfnamefont {G.~A.}\ \bibnamefont {Rochau}}, \bibinfo {author} {\bibfnamefont {G.~A.}\ \bibnamefont {Chandler}}, \bibinfo {author} {\bibfnamefont {J.~E.}\ \bibnamefont {Bailey}}, \bibinfo {author} {\bibfnamefont {J.~L.}\ \bibnamefont {Porter}}, \bibinfo {author} {\bibfnamefont {R.~D.}\ \bibnamefont {McBride}}, \bibinfo {author} {\bibfnamefont {D.~C.}\ \bibnamefont {Rovang}}, \bibinfo {author} {\bibfnamefont {M.~G.}\ \bibnamefont {Mazarakis}}, \bibinfo {author} {\bibfnamefont {E.~P.}\ \bibnamefont {Yu}}, \bibinfo
  {author} {\bibfnamefont {D.~C.}\ \bibnamefont {Lamppa}}, \bibinfo {author} {\bibfnamefont {K.~J.}\ \bibnamefont {Peterson}}, \bibinfo {author} {\bibfnamefont {C.}~\bibnamefont {Nakhleh}}, \bibinfo {author} {\bibfnamefont {S.~B.}\ \bibnamefont {Hansen}}, \bibinfo {author} {\bibfnamefont {A.~J.}\ \bibnamefont {Lopez}}, \bibinfo {author} {\bibfnamefont {M.~E.}\ \bibnamefont {Savage}}, \bibinfo {author} {\bibfnamefont {C.~A.}\ \bibnamefont {Jennings}}, \bibinfo {author} {\bibfnamefont {M.~R.}\ \bibnamefont {Martin}}, \bibinfo {author} {\bibfnamefont {R.~W.}\ \bibnamefont {Lemke}}, \bibinfo {author} {\bibfnamefont {B.~W.}\ \bibnamefont {Atherton}}, \bibinfo {author} {\bibfnamefont {I.~C.}\ \bibnamefont {Smith}}, \bibinfo {author} {\bibfnamefont {P.~K.}\ \bibnamefont {Rambo}}, \bibinfo {author} {\bibfnamefont {M.}~\bibnamefont {Jones}}, \bibinfo {author} {\bibfnamefont {M.~R.}\ \bibnamefont {Lopez}}, \bibinfo {author} {\bibfnamefont {P.~J.}\ \bibnamefont {Christenson}}, \bibinfo {author} {\bibfnamefont {M.~A.}\
  \bibnamefont {Sweeney}}, \bibinfo {author} {\bibfnamefont {B.}~\bibnamefont {Jones}}, \bibinfo {author} {\bibfnamefont {L.~A.}\ \bibnamefont {McPherson}}, \bibinfo {author} {\bibfnamefont {E.}~\bibnamefont {Harding}}, \bibinfo {author} {\bibfnamefont {M.~R.}\ \bibnamefont {Gomez}}, \bibinfo {author} {\bibfnamefont {P.~F.}\ \bibnamefont {Knapp}}, \bibinfo {author} {\bibfnamefont {T.~J.}\ \bibnamefont {Awe}}, \bibinfo {author} {\bibfnamefont {R.~J.}\ \bibnamefont {Leeper}}, \bibinfo {author} {\bibfnamefont {C.~L.}\ \bibnamefont {Ruiz}}, \bibinfo {author} {\bibfnamefont {G.~W.}\ \bibnamefont {Cooper}}, \bibinfo {author} {\bibfnamefont {K.~D.}\ \bibnamefont {Hahn}}, \bibinfo {author} {\bibfnamefont {J.}~\bibnamefont {McKenney}}, \bibinfo {author} {\bibfnamefont {A.~C.}\ \bibnamefont {Owen}}, \bibinfo {author} {\bibfnamefont {G.~R.}\ \bibnamefont {McKee}}, \bibinfo {author} {\bibfnamefont {G.~T.}\ \bibnamefont {Leifeste}}, \bibinfo {author} {\bibfnamefont {D.~J.}\ \bibnamefont {Ampleford}}, \bibinfo {author}
  {\bibfnamefont {E.~M.}\ \bibnamefont {Waisman}}, \bibinfo {author} {\bibfnamefont {A.}~\bibnamefont {Harvey-Thompson}}, \bibinfo {author} {\bibfnamefont {R.~J.}\ \bibnamefont {Kaye}}, \bibinfo {author} {\bibfnamefont {M.~H.}\ \bibnamefont {Hess}}, \bibinfo {author} {\bibfnamefont {S.~E.}\ \bibnamefont {Rosenthal}}, \ and\ \bibinfo {author} {\bibfnamefont {M.~K.}\ \bibnamefont {Matzen}},\ }\bibfield  {title} {\enquote {\bibinfo {title} {{Magnetically Driven Implosions for Inertial Confinement Fusion at Sandia National Laboratories}},}\ }\href {\doibase 10.1109/TPS.2012.2223488} {\bibfield  {journal} {\bibinfo  {journal} {IEEE Transactions on Plasma Science}\ }\textbf {\bibinfo {volume} {40}},\ \bibinfo {pages} {3222--3245} (\bibinfo {year} {2012})}\BibitemShut {NoStop}%
\bibitem [{\citenamefont {Gomez}\ \emph {et~al.}(2020)\citenamefont {Gomez}, \citenamefont {Slutz}, \citenamefont {Jennings}, \citenamefont {Ampleford}, \citenamefont {Weis}, \citenamefont {Myers}, \citenamefont {Yager-Elorriaga}, \citenamefont {Hahn}, \citenamefont {Hansen}, \citenamefont {Harding}, \citenamefont {Harvey-Thompson}, \citenamefont {Lamppa}, \citenamefont {Mangan}, \citenamefont {Knapp}, \citenamefont {Awe}, \citenamefont {Chandler}, \citenamefont {Cooper}, \citenamefont {Fein}, \citenamefont {Geissel}, \citenamefont {Glinsky}, \citenamefont {Lewis}, \citenamefont {Ruiz}, \citenamefont {Ruiz}, \citenamefont {Savage}, \citenamefont {Schmit}, \citenamefont {Smith}, \citenamefont {Styron}, \citenamefont {Porter}, \citenamefont {Jones}, \citenamefont {Mattsson}, \citenamefont {Peterson}, \citenamefont {Rochau},\ and\ \citenamefont {Sinars}}]{Gomez2020}%
  \BibitemOpen
  \bibfield  {author} {\bibinfo {author} {\bibfnamefont {M.~R.}\ \bibnamefont {Gomez}}, \bibinfo {author} {\bibfnamefont {S.~A.}\ \bibnamefont {Slutz}}, \bibinfo {author} {\bibfnamefont {C.~A.}\ \bibnamefont {Jennings}}, \bibinfo {author} {\bibfnamefont {D.~J.}\ \bibnamefont {Ampleford}}, \bibinfo {author} {\bibfnamefont {M.~R.}\ \bibnamefont {Weis}}, \bibinfo {author} {\bibfnamefont {C.~E.}\ \bibnamefont {Myers}}, \bibinfo {author} {\bibfnamefont {D.~A.}\ \bibnamefont {Yager-Elorriaga}}, \bibinfo {author} {\bibfnamefont {K.~D.}\ \bibnamefont {Hahn}}, \bibinfo {author} {\bibfnamefont {S.~B.}\ \bibnamefont {Hansen}}, \bibinfo {author} {\bibfnamefont {E.~C.}\ \bibnamefont {Harding}}, \bibinfo {author} {\bibfnamefont {A.~J.}\ \bibnamefont {Harvey-Thompson}}, \bibinfo {author} {\bibfnamefont {D.~C.}\ \bibnamefont {Lamppa}}, \bibinfo {author} {\bibfnamefont {M.}~\bibnamefont {Mangan}}, \bibinfo {author} {\bibfnamefont {P.~F.}\ \bibnamefont {Knapp}}, \bibinfo {author} {\bibfnamefont {T.~J.}\ \bibnamefont {Awe}},
  \bibinfo {author} {\bibfnamefont {G.~A.}\ \bibnamefont {Chandler}}, \bibinfo {author} {\bibfnamefont {G.~W.}\ \bibnamefont {Cooper}}, \bibinfo {author} {\bibfnamefont {J.~R.}\ \bibnamefont {Fein}}, \bibinfo {author} {\bibfnamefont {M.}~\bibnamefont {Geissel}}, \bibinfo {author} {\bibfnamefont {M.~E.}\ \bibnamefont {Glinsky}}, \bibinfo {author} {\bibfnamefont {W.~E.}\ \bibnamefont {Lewis}}, \bibinfo {author} {\bibfnamefont {C.~L.}\ \bibnamefont {Ruiz}}, \bibinfo {author} {\bibfnamefont {D.~E.}\ \bibnamefont {Ruiz}}, \bibinfo {author} {\bibfnamefont {M.~E.}\ \bibnamefont {Savage}}, \bibinfo {author} {\bibfnamefont {P.~F.}\ \bibnamefont {Schmit}}, \bibinfo {author} {\bibfnamefont {I.~C.}\ \bibnamefont {Smith}}, \bibinfo {author} {\bibfnamefont {J.~D.}\ \bibnamefont {Styron}}, \bibinfo {author} {\bibfnamefont {J.~L.}\ \bibnamefont {Porter}}, \bibinfo {author} {\bibfnamefont {B.}~\bibnamefont {Jones}}, \bibinfo {author} {\bibfnamefont {T.~R.}\ \bibnamefont {Mattsson}}, \bibinfo {author} {\bibfnamefont {K.~J.}\
  \bibnamefont {Peterson}}, \bibinfo {author} {\bibfnamefont {G.~A.}\ \bibnamefont {Rochau}}, \ and\ \bibinfo {author} {\bibfnamefont {D.~B.}\ \bibnamefont {Sinars}},\ }\bibfield  {title} {\enquote {\bibinfo {title} {Performance scaling in magnetized liner inertial fusion experiments},}\ }\href {\doibase 10.1103/PhysRevLett.125.155002} {\bibfield  {journal} {\bibinfo  {journal} {Phys. Rev. Lett.}\ }\textbf {\bibinfo {volume} {125}},\ \bibinfo {pages} {155002} (\bibinfo {year} {2020})}\BibitemShut {NoStop}%
\bibitem [{\citenamefont {Pollock}\ \emph {et~al.}(2023)\citenamefont {Pollock}, \citenamefont {Goyon}, \citenamefont {Sefkow}, \citenamefont {Glinsky}, \citenamefont {Peterson}, \citenamefont {Weis}, \citenamefont {Carroll}, \citenamefont {Fry}, \citenamefont {Piston}, \citenamefont {Harvey-Thompson}, \citenamefont {Hansen}, \citenamefont {Beckwith}, \citenamefont {Ampleford}, \citenamefont {Tubman}, \citenamefont {Strozzi}, \citenamefont {Ross},\ and\ \citenamefont {Moody}}]{MagLIFexp}%
  \BibitemOpen
  \bibfield  {author} {\bibinfo {author} {\bibfnamefont {B.~B.}\ \bibnamefont {Pollock}}, \bibinfo {author} {\bibfnamefont {C.}~\bibnamefont {Goyon}}, \bibinfo {author} {\bibfnamefont {A.~B.}\ \bibnamefont {Sefkow}}, \bibinfo {author} {\bibfnamefont {M.~E.}\ \bibnamefont {Glinsky}}, \bibinfo {author} {\bibfnamefont {K.~J.}\ \bibnamefont {Peterson}}, \bibinfo {author} {\bibfnamefont {M.~R.}\ \bibnamefont {Weis}}, \bibinfo {author} {\bibfnamefont {E.~G.}\ \bibnamefont {Carroll}}, \bibinfo {author} {\bibfnamefont {J.}~\bibnamefont {Fry}}, \bibinfo {author} {\bibfnamefont {K.}~\bibnamefont {Piston}}, \bibinfo {author} {\bibfnamefont {A.~J.}\ \bibnamefont {Harvey-Thompson}}, \bibinfo {author} {\bibfnamefont {S.~B.}\ \bibnamefont {Hansen}}, \bibinfo {author} {\bibfnamefont {K.}~\bibnamefont {Beckwith}}, \bibinfo {author} {\bibfnamefont {D.~J.}\ \bibnamefont {Ampleford}}, \bibinfo {author} {\bibfnamefont {E.~R.}\ \bibnamefont {Tubman}}, \bibinfo {author} {\bibfnamefont {D.~J.}\ \bibnamefont {Strozzi}}, \bibinfo
  {author} {\bibfnamefont {J.~S.}\ \bibnamefont {Ross}}, \ and\ \bibinfo {author} {\bibfnamefont {J.~D.}\ \bibnamefont {Moody}},\ }\bibfield  {title} {\enquote {\bibinfo {title} {{Experimental demonstration of >20 kJ laser energy coupling in 1-cm hydrocarbon-filled gas pipe targets via inverse Bremsstrahlung absorption with applications to MagLIF}},}\ }\href {\doibase 10.1063/5.0120916} {\bibfield  {journal} {\bibinfo  {journal} {Physics of Plasmas}\ }\textbf {\bibinfo {volume} {30}},\ \bibinfo {pages} {022711} (\bibinfo {year} {2023})}\BibitemShut {NoStop}%
\bibitem [{\citenamefont {Weis}\ \emph {et~al.}(2025)\citenamefont {Weis}, \citenamefont {Ruiz}, \citenamefont {Gomez}, \citenamefont {Harvey-Thompson}, \citenamefont {Jennings}, \citenamefont {Yager-Elorriaga}, \citenamefont {Lewis}, \citenamefont {Slutz}, \citenamefont {Shulenburger}, \citenamefont {Ampleford}, \citenamefont {Beckwith},\ and\ \citenamefont {Koning}}]{Weis2025}%
  \BibitemOpen
  \bibfield  {author} {\bibinfo {author} {\bibfnamefont {M.~R.}\ \bibnamefont {Weis}}, \bibinfo {author} {\bibfnamefont {D.~E.}\ \bibnamefont {Ruiz}}, \bibinfo {author} {\bibfnamefont {M.~R.}\ \bibnamefont {Gomez}}, \bibinfo {author} {\bibfnamefont {A.~J.}\ \bibnamefont {Harvey-Thompson}}, \bibinfo {author} {\bibfnamefont {C.~A.}\ \bibnamefont {Jennings}}, \bibinfo {author} {\bibfnamefont {D.~A.}\ \bibnamefont {Yager-Elorriaga}}, \bibinfo {author} {\bibfnamefont {W.~E.}\ \bibnamefont {Lewis}}, \bibinfo {author} {\bibfnamefont {S.~A.}\ \bibnamefont {Slutz}}, \bibinfo {author} {\bibfnamefont {L.}~\bibnamefont {Shulenburger}}, \bibinfo {author} {\bibfnamefont {D.~J.}\ \bibnamefont {Ampleford}}, \bibinfo {author} {\bibfnamefont {K.}~\bibnamefont {Beckwith}}, \ and\ \bibinfo {author} {\bibfnamefont {J.~M.}\ \bibnamefont {Koning}},\ }\bibfield  {title} {\enquote {\bibinfo {title} {Assessing the performance of maglif with 3d mhd simulations},}\ }\href {\doibase 10.1063/5.0244304} {\bibfield  {journal} {\bibinfo
  {journal} {Physics of Plasmas}\ }\textbf {\bibinfo {volume} {32}},\ \bibinfo {pages} {022708} (\bibinfo {year} {2025})},\ \Eprint {http://arxiv.org/abs/https://pubs.aip.org/aip/pop/article-pdf/doi/10.1063/5.0244304/20404620/022708\_1\_5.0244304.pdf} {https://pubs.aip.org/aip/pop/article-pdf/doi/10.1063/5.0244304/20404620/022708\_1\_5.0244304.pdf} \BibitemShut {NoStop}%
\bibitem [{\citenamefont {Shvarts}\ \emph {et~al.}(1981)\citenamefont {Shvarts}, \citenamefont {Delettrez}, \citenamefont {McCrory},\ and\ \citenamefont {Verdon}}]{Shvarts1981}%
  \BibitemOpen
  \bibfield  {author} {\bibinfo {author} {\bibfnamefont {D.}~\bibnamefont {Shvarts}}, \bibinfo {author} {\bibfnamefont {J.}~\bibnamefont {Delettrez}}, \bibinfo {author} {\bibfnamefont {R.~L.}\ \bibnamefont {McCrory}}, \ and\ \bibinfo {author} {\bibfnamefont {C.~P.}\ \bibnamefont {Verdon}},\ }\bibfield  {title} {\enquote {\bibinfo {title} {{Self-Consistent Reduction of the Spitzer-H\"arm Electron Thermal Heat Flux in Steep Temperature Gradients in Laser-Produced Plasmas}},}\ }\href {\doibase 10.1103/PhysRevLett.47.247} {\bibfield  {journal} {\bibinfo  {journal} {Phys. Rev. Lett.}\ }\textbf {\bibinfo {volume} {47}},\ \bibinfo {pages} {247--250} (\bibinfo {year} {1981})}\BibitemShut {NoStop}%
\bibitem [{\citenamefont {Malone}, \citenamefont {McCrory},\ and\ \citenamefont {Morse}(1975)}]{fluxLimiter}%
  \BibitemOpen
  \bibfield  {author} {\bibinfo {author} {\bibfnamefont {R.~C.}\ \bibnamefont {Malone}}, \bibinfo {author} {\bibfnamefont {R.~L.}\ \bibnamefont {McCrory}}, \ and\ \bibinfo {author} {\bibfnamefont {R.~L.}\ \bibnamefont {Morse}},\ }\bibfield  {title} {\enquote {\bibinfo {title} {Indications of strongly flux-limited electron thermal conduction in laser-target experiments},}\ }\href@noop {} {\bibfield  {journal} {\bibinfo  {journal} {Phys. Rev. Lett.}\ }\textbf {\bibinfo {volume} {34}} (\bibinfo {year} {1975})}\BibitemShut {NoStop}%
\bibitem [{\citenamefont {Goncharov}\ \emph {et~al.}(2006{\natexlab{a}})\citenamefont {Goncharov}, \citenamefont {Gotchev}, \citenamefont {Vianello}, \citenamefont {Boehly}, \citenamefont {Knauer}, \citenamefont {McKenty}, \citenamefont {Radha}, \citenamefont {Regan}, \citenamefont {Sangster}, \citenamefont {Skupsky}, \citenamefont {Smalyuk}, \citenamefont {Betti}, \citenamefont {McCrory}, \citenamefont {Meyerhofer},\ and\ \citenamefont {Cherfils-Clérouin}}]{Goncharov2006}%
  \BibitemOpen
  \bibfield  {author} {\bibinfo {author} {\bibfnamefont {V.~N.}\ \bibnamefont {Goncharov}}, \bibinfo {author} {\bibfnamefont {O.~V.}\ \bibnamefont {Gotchev}}, \bibinfo {author} {\bibfnamefont {E.}~\bibnamefont {Vianello}}, \bibinfo {author} {\bibfnamefont {T.~R.}\ \bibnamefont {Boehly}}, \bibinfo {author} {\bibfnamefont {J.~P.}\ \bibnamefont {Knauer}}, \bibinfo {author} {\bibfnamefont {P.~W.}\ \bibnamefont {McKenty}}, \bibinfo {author} {\bibfnamefont {P.~B.}\ \bibnamefont {Radha}}, \bibinfo {author} {\bibfnamefont {S.~P.}\ \bibnamefont {Regan}}, \bibinfo {author} {\bibfnamefont {T.~C.}\ \bibnamefont {Sangster}}, \bibinfo {author} {\bibfnamefont {S.}~\bibnamefont {Skupsky}}, \bibinfo {author} {\bibfnamefont {V.~A.}\ \bibnamefont {Smalyuk}}, \bibinfo {author} {\bibfnamefont {R.}~\bibnamefont {Betti}}, \bibinfo {author} {\bibfnamefont {R.~L.}\ \bibnamefont {McCrory}}, \bibinfo {author} {\bibfnamefont {D.~D.}\ \bibnamefont {Meyerhofer}}, \ and\ \bibinfo {author} {\bibfnamefont {C.}~\bibnamefont
  {Cherfils-Clérouin}},\ }\bibfield  {title} {\enquote {\bibinfo {title} {{Early stage of implosion in inertial confinement fusion: Shock timing and perturbation evolution}},}\ }\href {\doibase 10.1063/1.2162803} {\bibfield  {journal} {\bibinfo  {journal} {Physics of Plasmas}\ }\textbf {\bibinfo {volume} {13}},\ \bibinfo {pages} {012702} (\bibinfo {year} {2006}{\natexlab{a}})},\ \Eprint {http://arxiv.org/abs/https://pubs.aip.org/aip/pop/article-pdf/doi/10.1063/1.2162803/15984853/012702\_1\_online.pdf} {https://pubs.aip.org/aip/pop/article-pdf/doi/10.1063/1.2162803/15984853/012702\_1\_online.pdf} \BibitemShut {NoStop}%
\bibitem [{\citenamefont {Goncharov}\ \emph {et~al.}(2006{\natexlab{b}})\citenamefont {Goncharov}, \citenamefont {Gotchev}, \citenamefont {McCrory}, \citenamefont {McKenty}, \citenamefont {Meyerhofer}, \citenamefont {Sangster}, \citenamefont {Skupsky},\ and\ \citenamefont {Cherfils-Clerouin}}]{Goncharov2005}%
  \BibitemOpen
  \bibfield  {author} {\bibinfo {author} {\bibfnamefont {V.}~\bibnamefont {Goncharov}}, \bibinfo {author} {\bibfnamefont {O.}~\bibnamefont {Gotchev}}, \bibinfo {author} {\bibfnamefont {R.}~\bibnamefont {McCrory}}, \bibinfo {author} {\bibfnamefont {P.}~\bibnamefont {McKenty}}, \bibinfo {author} {\bibfnamefont {D.}~\bibnamefont {Meyerhofer}}, \bibinfo {author} {\bibfnamefont {T.}~\bibnamefont {Sangster}}, \bibinfo {author} {\bibfnamefont {S.}~\bibnamefont {Skupsky}}, \ and\ \bibinfo {author} {\bibfnamefont {C.}~\bibnamefont {Cherfils-Clerouin}},\ }\bibfield  {title} {\enquote {\bibinfo {title} {{Ablative Richtmyer-$\!$-Meshkov instability: Theory and experimental results}},}\ }\href@noop {} {\bibfield  {journal} {\bibinfo  {journal} {J. Phys. IV France}\ }\textbf {\bibinfo {volume} {133}},\ \bibinfo {pages} {123--127} (\bibinfo {year} {2006}{\natexlab{b}})}\BibitemShut {NoStop}%
\bibitem [{\citenamefont {Boehly}\ \emph {et~al.}(2011)\citenamefont {Boehly}, \citenamefont {Goncharov}, \citenamefont {Seka}, \citenamefont {Hu}, \citenamefont {Marozas}, \citenamefont {Meyerhofer}, \citenamefont {Celliers}, \citenamefont {Hicks}, \citenamefont {Barrios}, \citenamefont {Fratanduono},\ and\ \citenamefont {Collins}}]{Boehly2011}%
  \BibitemOpen
  \bibfield  {author} {\bibinfo {author} {\bibfnamefont {T.~R.}\ \bibnamefont {Boehly}}, \bibinfo {author} {\bibfnamefont {V.~N.}\ \bibnamefont {Goncharov}}, \bibinfo {author} {\bibfnamefont {W.}~\bibnamefont {Seka}}, \bibinfo {author} {\bibfnamefont {S.~X.}\ \bibnamefont {Hu}}, \bibinfo {author} {\bibfnamefont {J.~A.}\ \bibnamefont {Marozas}}, \bibinfo {author} {\bibfnamefont {D.~D.}\ \bibnamefont {Meyerhofer}}, \bibinfo {author} {\bibfnamefont {P.~M.}\ \bibnamefont {Celliers}}, \bibinfo {author} {\bibfnamefont {D.~G.}\ \bibnamefont {Hicks}}, \bibinfo {author} {\bibfnamefont {M.~A.}\ \bibnamefont {Barrios}}, \bibinfo {author} {\bibfnamefont {D.}~\bibnamefont {Fratanduono}}, \ and\ \bibinfo {author} {\bibfnamefont {G.~W.}\ \bibnamefont {Collins}},\ }\bibfield  {title} {\enquote {\bibinfo {title} {{Multiple spherically converging shock waves in liquid deuterium}},}\ }\href {\doibase 10.1063/1.3640805} {\bibfield  {journal} {\bibinfo  {journal} {Physics of Plasmas}\ }\textbf {\bibinfo {volume} {18}},\ \bibinfo
  {pages} {092706} (\bibinfo {year} {2011})},\ \Eprint {http://arxiv.org/abs/https://pubs.aip.org/aip/pop/article-pdf/doi/10.1063/1.3640805/13686549/092706\_1\_online.pdf} {https://pubs.aip.org/aip/pop/article-pdf/doi/10.1063/1.3640805/13686549/092706\_1\_online.pdf} \BibitemShut {NoStop}%
\bibitem [{\citenamefont {Spitzer}\ and\ \citenamefont {H\"arm}(1953)}]{Spitzer}%
  \BibitemOpen
  \bibfield  {author} {\bibinfo {author} {\bibfnamefont {L.}~\bibnamefont {Spitzer}}\ and\ \bibinfo {author} {\bibfnamefont {R.}~\bibnamefont {H\"arm}},\ }\bibfield  {title} {\enquote {\bibinfo {title} {Transport phenomena in a completely ionized gas},}\ }\href {\doibase 10.1103/PhysRev.89.977} {\bibfield  {journal} {\bibinfo  {journal} {Phys. Rev.}\ }\textbf {\bibinfo {volume} {89}},\ \bibinfo {pages} {977--981} (\bibinfo {year} {1953})}\BibitemShut {NoStop}%
\bibitem [{\citenamefont {{Braginskii}}(1965)}]{Braginskii}%
  \BibitemOpen
  \bibfield  {author} {\bibinfo {author} {\bibfnamefont {S.~I.}\ \bibnamefont {{Braginskii}}},\ }\bibfield  {title} {\enquote {\bibinfo {title} {{Transport Processes in a Plasma}},}\ }\href@noop {} {\bibfield  {journal} {\bibinfo  {journal} {Reviews of Plasma Physics}\ }\textbf {\bibinfo {volume} {1}},\ \bibinfo {pages} {205} (\bibinfo {year} {1965})}\BibitemShut {NoStop}%
\bibitem [{\citenamefont {Davies}\ \emph {et~al.}(2015)\citenamefont {Davies}, \citenamefont {Betti}, \citenamefont {Chang},\ and\ \citenamefont {Fiksel}}]{Davies2015}%
  \BibitemOpen
  \bibfield  {author} {\bibinfo {author} {\bibfnamefont {J.~R.}\ \bibnamefont {Davies}}, \bibinfo {author} {\bibfnamefont {R.}~\bibnamefont {Betti}}, \bibinfo {author} {\bibfnamefont {P.-Y.}\ \bibnamefont {Chang}}, \ and\ \bibinfo {author} {\bibfnamefont {G.}~\bibnamefont {Fiksel}},\ }\bibfield  {title} {\enquote {\bibinfo {title} {{The importance of electrothermal terms in Ohm's law for magnetized spherical implosions}},}\ }\href {\doibase 10.1063/1.4935286} {\bibfield  {journal} {\bibinfo  {journal} {Physics of Plasmas}\ }\textbf {\bibinfo {volume} {22}},\ \bibinfo {pages} {112703} (\bibinfo {year} {2015})},\ \Eprint {http://arxiv.org/abs/https://pubs.aip.org/aip/pop/article-pdf/doi/10.1063/1.4935286/14870934/112703\_1\_online.pdf} {https://pubs.aip.org/aip/pop/article-pdf/doi/10.1063/1.4935286/14870934/112703\_1\_online.pdf} \BibitemShut {NoStop}%
\bibitem [{\citenamefont {Bell}, \citenamefont {Evans},\ and\ \citenamefont {Nicholas}(1981)}]{Bell1981}%
  \BibitemOpen
  \bibfield  {author} {\bibinfo {author} {\bibfnamefont {A.~R.}\ \bibnamefont {Bell}}, \bibinfo {author} {\bibfnamefont {R.~G.}\ \bibnamefont {Evans}}, \ and\ \bibinfo {author} {\bibfnamefont {D.~J.}\ \bibnamefont {Nicholas}},\ }\bibfield  {title} {\enquote {\bibinfo {title} {Elecron energy transport in steep temperature gradients in laser-produced plasmas},}\ }\href {\doibase 10.1103/PhysRevLett.46.243} {\bibfield  {journal} {\bibinfo  {journal} {Phys. Rev. Lett.}\ }\textbf {\bibinfo {volume} {46}},\ \bibinfo {pages} {243--246} (\bibinfo {year} {1981})}\BibitemShut {NoStop}%
\bibitem [{\citenamefont {Johnston}(1960)}]{CartesianExpansion}%
  \BibitemOpen
  \bibfield  {author} {\bibinfo {author} {\bibfnamefont {T.~W.}\ \bibnamefont {Johnston}},\ }\bibfield  {title} {\enquote {\bibinfo {title} {{Cartesian Tensor Scalar Product and Spherical Harmonic Expansions in Boltzmann's Equation}},}\ }\href {\doibase 10.1103/PhysRev.120.1103} {\bibfield  {journal} {\bibinfo  {journal} {Phys. Rev.}\ }\textbf {\bibinfo {volume} {120}},\ \bibinfo {pages} {1103--1111} (\bibinfo {year} {1960})}\BibitemShut {NoStop}%
\bibitem [{\citenamefont {Epperlein}\ and\ \citenamefont {Short}(1991)}]{EpperleinShort}%
  \BibitemOpen
  \bibfield  {author} {\bibinfo {author} {\bibfnamefont {E.~M.}\ \bibnamefont {Epperlein}}\ and\ \bibinfo {author} {\bibfnamefont {R.~W.}\ \bibnamefont {Short}},\ }\bibfield  {title} {\enquote {\bibinfo {title} {{A practical nonlocal model for electron heat transport in laser plasmas}},}\ }\href {\doibase 10.1063/1.859789} {\bibfield  {journal} {\bibinfo  {journal} {Physics of Fluids B: Plasma Physics}\ }\textbf {\bibinfo {volume} {3}},\ \bibinfo {pages} {3092--3098} (\bibinfo {year} {1991})},\ \Eprint {http://arxiv.org/abs/https://pubs.aip.org/aip/pfb/article-pdf/3/11/3092/12317455/3092\_1\_online.pdf} {https://pubs.aip.org/aip/pfb/article-pdf/3/11/3092/12317455/3092\_1\_online.pdf} \BibitemShut {NoStop}%
\bibitem [{\citenamefont {Davies}(2023)}]{Davies}%
  \BibitemOpen
  \bibfield  {author} {\bibinfo {author} {\bibfnamefont {J.~R.}\ \bibnamefont {Davies}},\ }\bibfield  {title} {\enquote {\bibinfo {title} {{Nonlocal suppression of Biermann battery magnetic-field generation for arbitrary atomic numbers and magnetization}},}\ }\href {\doibase 10.1063/5.0152530} {\bibfield  {journal} {\bibinfo  {journal} {Physics of Plasmas}\ }\textbf {\bibinfo {volume} {30}},\ \bibinfo {pages} {072701} (\bibinfo {year} {2023})}\BibitemShut {NoStop}%
\bibitem [{\citenamefont {Sherlock}, \citenamefont {Brodrick},\ and\ \citenamefont {Ridgers}(2017)}]{K2ref}%
  \BibitemOpen
  \bibfield  {author} {\bibinfo {author} {\bibfnamefont {M.}~\bibnamefont {Sherlock}}, \bibinfo {author} {\bibfnamefont {J.~P.}\ \bibnamefont {Brodrick}}, \ and\ \bibinfo {author} {\bibfnamefont {C.~P.}\ \bibnamefont {Ridgers}},\ }\bibfield  {title} {\enquote {\bibinfo {title} {{A comparison of non-local electron transport models for laser-plasmas relevant to inertial confinement fusion}},}\ }\href {\doibase 10.1063/1.4986095} {\bibfield  {journal} {\bibinfo  {journal} {Physics of Plasmas}\ }\textbf {\bibinfo {volume} {24}},\ \bibinfo {pages} {082706} (\bibinfo {year} {2017})}\BibitemShut {NoStop}%
\bibitem [{\citenamefont {Marinak}\ \emph {et~al.}(2001)\citenamefont {Marinak}, \citenamefont {Kerbel}, \citenamefont {Gentile}, \citenamefont {Jones}, \citenamefont {Munro}, \citenamefont {Pollaine}, \citenamefont {Dittrich},\ and\ \citenamefont {Haan}}]{HydraRef}%
  \BibitemOpen
  \bibfield  {author} {\bibinfo {author} {\bibfnamefont {M.~M.}\ \bibnamefont {Marinak}}, \bibinfo {author} {\bibfnamefont {G.~D.}\ \bibnamefont {Kerbel}}, \bibinfo {author} {\bibfnamefont {N.~A.}\ \bibnamefont {Gentile}}, \bibinfo {author} {\bibfnamefont {O.}~\bibnamefont {Jones}}, \bibinfo {author} {\bibfnamefont {D.}~\bibnamefont {Munro}}, \bibinfo {author} {\bibfnamefont {S.}~\bibnamefont {Pollaine}}, \bibinfo {author} {\bibfnamefont {T.~R.}\ \bibnamefont {Dittrich}}, \ and\ \bibinfo {author} {\bibfnamefont {S.~W.}\ \bibnamefont {Haan}},\ }\bibfield  {title} {\enquote {\bibinfo {title} {{Three-dimensional HYDRA simulations of National Ignition Facility targets}},}\ }\href {\doibase 10.1063/1.1356740} {\bibfield  {journal} {\bibinfo  {journal} {Physics of Plasmas}\ }\textbf {\bibinfo {volume} {8}},\ \bibinfo {pages} {2275--2280} (\bibinfo {year} {2001})},\ \Eprint {http://arxiv.org/abs/https://pubs.aip.org/aip/pop/article-pdf/8/5/2275/19187439/2275\_1\_online.pdf}
  {https://pubs.aip.org/aip/pop/article-pdf/8/5/2275/19187439/2275\_1\_online.pdf} \BibitemShut {NoStop}%
\bibitem [{\citenamefont {Schurtz}, \citenamefont {Nicolaï},\ and\ \citenamefont {Busquet}(2000)}]{SNB}%
  \BibitemOpen
  \bibfield  {author} {\bibinfo {author} {\bibfnamefont {G.~P.}\ \bibnamefont {Schurtz}}, \bibinfo {author} {\bibfnamefont {P.~D.}\ \bibnamefont {Nicolaï}}, \ and\ \bibinfo {author} {\bibfnamefont {M.}~\bibnamefont {Busquet}},\ }\bibfield  {title} {\enquote {\bibinfo {title} {{A nonlocal electron conduction model for multidimensional radiation hydrodynamics codes}},}\ }\href {\doibase 10.1063/1.1289512} {\bibfield  {journal} {\bibinfo  {journal} {Physics of Plasmas}\ }\textbf {\bibinfo {volume} {7}},\ \bibinfo {pages} {4238--4249} (\bibinfo {year} {2000})}\BibitemShut {NoStop}%
\bibitem [{\citenamefont {Robinson}, \citenamefont {Bell},\ and\ \citenamefont {Kingham}(2006)}]{Kalos}%
  \BibitemOpen
  \bibfield  {author} {\bibinfo {author} {\bibfnamefont {A.~P.~L.}\ \bibnamefont {Robinson}}, \bibinfo {author} {\bibfnamefont {A.~R.}\ \bibnamefont {Bell}}, \ and\ \bibinfo {author} {\bibfnamefont {R.~J.}\ \bibnamefont {Kingham}},\ }\bibfield  {title} {\enquote {\bibinfo {title} {Fast electron transport and ionization in a target irradiated by a high power laser},}\ }\href {\doibase 10.1088/0741-3335/48/8/002} {\bibfield  {journal} {\bibinfo  {journal} {Plasma Physics and Controlled Fusion}\ }\textbf {\bibinfo {volume} {48}},\ \bibinfo {pages} {1063} (\bibinfo {year} {2006})}\BibitemShut {NoStop}%
\bibitem [{\citenamefont {Tzoufras}\ \emph {et~al.}(2011)\citenamefont {Tzoufras}, \citenamefont {Bell}, \citenamefont {Norreys},\ and\ \citenamefont {Tsung}}]{TzoufrasOSHUN}%
  \BibitemOpen
  \bibfield  {author} {\bibinfo {author} {\bibfnamefont {M.}~\bibnamefont {Tzoufras}}, \bibinfo {author} {\bibfnamefont {A.}~\bibnamefont {Bell}}, \bibinfo {author} {\bibfnamefont {P.}~\bibnamefont {Norreys}}, \ and\ \bibinfo {author} {\bibfnamefont {F.}~\bibnamefont {Tsung}},\ }\bibfield  {title} {\enquote {\bibinfo {title} {{A Vlasov–Fokker–Planck code for high energy density physics}},}\ }\href {\doibase https://doi.org/10.1016/j.jcp.2011.04.034} {\bibfield  {journal} {\bibinfo  {journal} {Journal of Computational Physics}\ }\textbf {\bibinfo {volume} {230}},\ \bibinfo {pages} {6475--6494} (\bibinfo {year} {2011})}\BibitemShut {NoStop}%
\bibitem [{\citenamefont {Rosenbluth}, \citenamefont {MacDonald},\ and\ \citenamefont {Judd}(1957)}]{Rosenbluth}%
  \BibitemOpen
  \bibfield  {author} {\bibinfo {author} {\bibfnamefont {M.~N.}\ \bibnamefont {Rosenbluth}}, \bibinfo {author} {\bibfnamefont {W.~M.}\ \bibnamefont {MacDonald}}, \ and\ \bibinfo {author} {\bibfnamefont {D.~L.}\ \bibnamefont {Judd}},\ }\bibfield  {title} {\enquote {\bibinfo {title} {{Fokker-Planck Equation for an Inverse-Square Force}},}\ }\href {\doibase 10.1103/PhysRev.107.1} {\bibfield  {journal} {\bibinfo  {journal} {Phys. Rev.}\ }\textbf {\bibinfo {volume} {107}},\ \bibinfo {pages} {1--6} (\bibinfo {year} {1957})}\BibitemShut {NoStop}%
\bibitem [{\citenamefont {Huba}(2004)}]{Huba2004NRLPF}%
  \BibitemOpen
  \bibfield  {author} {\bibinfo {author} {\bibfnamefont {J.~D.}\ \bibnamefont {Huba}},\ }\bibfield  {title} {\enquote {\bibinfo {title} {{NRL: Plasma Formulary}},}\ \ }(\bibinfo {year} {2004})\BibitemShut {NoStop}%
\bibitem [{\citenamefont {Epperlein}\ and\ \citenamefont {Haines}(1986)}]{EpperleinHaines}%
  \BibitemOpen
  \bibfield  {author} {\bibinfo {author} {\bibfnamefont {E.~M.}\ \bibnamefont {Epperlein}}\ and\ \bibinfo {author} {\bibfnamefont {M.~G.}\ \bibnamefont {Haines}},\ }\bibfield  {title} {\enquote {\bibinfo {title} {{Plasma transport coefficients in a magnetic field by direct numerical solution of the Fokker–Planck equation}},}\ }\href {\doibase 10.1063/1.865901} {\bibfield  {journal} {\bibinfo  {journal} {The Physics of Fluids}\ }\textbf {\bibinfo {volume} {29}},\ \bibinfo {pages} {1029--1041} (\bibinfo {year} {1986})},\ \Eprint {http://arxiv.org/abs/https://pubs.aip.org/aip/pfl/article-pdf/29/4/1029/12593728/1029\_1\_online.pdf} {https://pubs.aip.org/aip/pfl/article-pdf/29/4/1029/12593728/1029\_1\_online.pdf} \BibitemShut {NoStop}%
\bibitem [{\citenamefont {Bell}(1985)}]{Bell1985}%
  \BibitemOpen
  \bibfield  {author} {\bibinfo {author} {\bibfnamefont {A.~R.}\ \bibnamefont {Bell}},\ }\bibfield  {title} {\enquote {\bibinfo {title} {{Non‐Spitzer heat flow in a steadily ablating laser‐produced plasma}},}\ }\href {\doibase 10.1063/1.865378} {\bibfield  {journal} {\bibinfo  {journal} {The Physics of Fluids}\ }\textbf {\bibinfo {volume} {28}},\ \bibinfo {pages} {2007--2014} (\bibinfo {year} {1985})},\ \Eprint {http://arxiv.org/abs/https://pubs.aip.org/aip/pfl/article-pdf/28/6/2007/12702622/2007\_1\_online.pdf} {https://pubs.aip.org/aip/pfl/article-pdf/28/6/2007/12702622/2007\_1\_online.pdf} \BibitemShut {NoStop}%
\bibitem [{\citenamefont {Rosen}\ \emph {et~al.}(2011)\citenamefont {Rosen}, \citenamefont {Scott}, \citenamefont {Hinkel}, \citenamefont {Williams}, \citenamefont {Callahan}, \citenamefont {Town}, \citenamefont {Divol}, \citenamefont {Michel}, \citenamefont {Kruer}, \citenamefont {Suter}, \citenamefont {London}, \citenamefont {Harte},\ and\ \citenamefont {Zimmerman}}]{Rosen2011}%
  \BibitemOpen
  \bibfield  {author} {\bibinfo {author} {\bibfnamefont {M.}~\bibnamefont {Rosen}}, \bibinfo {author} {\bibfnamefont {H.}~\bibnamefont {Scott}}, \bibinfo {author} {\bibfnamefont {D.}~\bibnamefont {Hinkel}}, \bibinfo {author} {\bibfnamefont {E.}~\bibnamefont {Williams}}, \bibinfo {author} {\bibfnamefont {D.}~\bibnamefont {Callahan}}, \bibinfo {author} {\bibfnamefont {R.}~\bibnamefont {Town}}, \bibinfo {author} {\bibfnamefont {L.}~\bibnamefont {Divol}}, \bibinfo {author} {\bibfnamefont {P.}~\bibnamefont {Michel}}, \bibinfo {author} {\bibfnamefont {W.}~\bibnamefont {Kruer}}, \bibinfo {author} {\bibfnamefont {L.}~\bibnamefont {Suter}}, \bibinfo {author} {\bibfnamefont {R.}~\bibnamefont {London}}, \bibinfo {author} {\bibfnamefont {J.}~\bibnamefont {Harte}}, \ and\ \bibinfo {author} {\bibfnamefont {G.}~\bibnamefont {Zimmerman}},\ }\bibfield  {title} {\enquote {\bibinfo {title} {{The role of a detailed configuration accounting (DCA) atomic physics package in explaining the energy balance in ignition-scale
  hohlraums}},}\ }\href {\doibase https://doi.org/10.1016/j.hedp.2011.03.008} {\bibfield  {journal} {\bibinfo  {journal} {High Energy Density Physics}\ }\textbf {\bibinfo {volume} {7}},\ \bibinfo {pages} {180--190} (\bibinfo {year} {2011})}\BibitemShut {NoStop}%
\bibitem [{\citenamefont {Walsh}\ and\ \citenamefont {Sherlock}(2024)}]{Walsh2024}%
  \BibitemOpen
  \bibfield  {author} {\bibinfo {author} {\bibfnamefont {C.~A.}\ \bibnamefont {Walsh}}\ and\ \bibinfo {author} {\bibfnamefont {M.}~\bibnamefont {Sherlock}},\ }\bibfield  {title} {\enquote {\bibinfo {title} {{Kinetic corrections to heat-flow and Nernst advection for laser heated plasmas}},}\ }\href {\doibase 10.1063/5.0225592} {\bibfield  {journal} {\bibinfo  {journal} {Physics of Plasmas}\ }\textbf {\bibinfo {volume} {31}},\ \bibinfo {pages} {102704} (\bibinfo {year} {2024})},\ \Eprint {http://arxiv.org/abs/https://pubs.aip.org/aip/pop/article-pdf/doi/10.1063/5.0225592/20204656/102704\_1\_5.0225592.pdf} {https://pubs.aip.org/aip/pop/article-pdf/doi/10.1063/5.0225592/20204656/102704\_1\_5.0225592.pdf} \BibitemShut {NoStop}%
\bibitem [{\citenamefont {Bell}(1983{\natexlab{a}})}]{Bell1983}%
  \BibitemOpen
  \bibfield  {author} {\bibinfo {author} {\bibfnamefont {A.~R.}\ \bibnamefont {Bell}},\ }\bibfield  {title} {\enquote {\bibinfo {title} {{Electron energy transport in ion waves and its relevance to laser‐produced plasmas}},}\ }\href {\doibase 10.1063/1.864018} {\bibfield  {journal} {\bibinfo  {journal} {The Physics of Fluids}\ }\textbf {\bibinfo {volume} {26}},\ \bibinfo {pages} {279--284} (\bibinfo {year} {1983}{\natexlab{a}})},\ \Eprint {http://arxiv.org/abs/https://pubs.aip.org/aip/pfl/article-pdf/26/1/279/12262372/279\_1\_online.pdf} {https://pubs.aip.org/aip/pfl/article-pdf/26/1/279/12262372/279\_1\_online.pdf} \BibitemShut {NoStop}%
\bibitem [{\citenamefont {Luciani}, \citenamefont {Mora},\ and\ \citenamefont {Virmont}(1983)}]{LMV}%
  \BibitemOpen
  \bibfield  {author} {\bibinfo {author} {\bibfnamefont {J.~F.}\ \bibnamefont {Luciani}}, \bibinfo {author} {\bibfnamefont {P.}~\bibnamefont {Mora}}, \ and\ \bibinfo {author} {\bibfnamefont {J.}~\bibnamefont {Virmont}},\ }\bibfield  {title} {\enquote {\bibinfo {title} {Nonlocal heat transport due to steep temperature gradients},}\ }\href {\doibase 10.1103/PhysRevLett.51.1664} {\bibfield  {journal} {\bibinfo  {journal} {Phys. Rev. Lett.}\ }\textbf {\bibinfo {volume} {51}},\ \bibinfo {pages} {1664--1667} (\bibinfo {year} {1983})}\BibitemShut {NoStop}%
\bibitem [{\citenamefont {Manheimer}, \citenamefont {Colombant},\ and\ \citenamefont {Goncharov}(2008)}]{CMG}%
  \BibitemOpen
  \bibfield  {author} {\bibinfo {author} {\bibfnamefont {W.}~\bibnamefont {Manheimer}}, \bibinfo {author} {\bibfnamefont {D.}~\bibnamefont {Colombant}}, \ and\ \bibinfo {author} {\bibfnamefont {V.}~\bibnamefont {Goncharov}},\ }\bibfield  {title} {\enquote {\bibinfo {title} {{The development of a Krook model for nonlocal transport in laser produced plasmas. I. Basic theory}},}\ }\href {\doibase 10.1063/1.2963078} {\bibfield  {journal} {\bibinfo  {journal} {Physics of Plasmas}\ }\textbf {\bibinfo {volume} {15}},\ \bibinfo {pages} {083103} (\bibinfo {year} {2008})},\ \Eprint {http://arxiv.org/abs/https://pubs.aip.org/aip/pop/article-pdf/doi/10.1063/1.2963078/15625206/083103\_1\_online.pdf} {https://pubs.aip.org/aip/pop/article-pdf/doi/10.1063/1.2963078/15625206/083103\_1\_online.pdf} \BibitemShut {NoStop}%
\bibitem [{\citenamefont {Brodrick}\ \emph {et~al.}(2017)\citenamefont {Brodrick}, \citenamefont {Kingham}, \citenamefont {Marinak}, \citenamefont {Patel}, \citenamefont {Chankin}, \citenamefont {Omotani}, \citenamefont {Umansky}, \citenamefont {Del~Sorbo}, \citenamefont {Dudson}, \citenamefont {Parker}, \citenamefont {Kerbel}, \citenamefont {Sherlock},\ and\ \citenamefont {Ridgers}}]{BrodrickCorrection}%
  \BibitemOpen
  \bibfield  {author} {\bibinfo {author} {\bibfnamefont {J.~P.}\ \bibnamefont {Brodrick}}, \bibinfo {author} {\bibfnamefont {R.~J.}\ \bibnamefont {Kingham}}, \bibinfo {author} {\bibfnamefont {M.~M.}\ \bibnamefont {Marinak}}, \bibinfo {author} {\bibfnamefont {M.~V.}\ \bibnamefont {Patel}}, \bibinfo {author} {\bibfnamefont {A.~V.}\ \bibnamefont {Chankin}}, \bibinfo {author} {\bibfnamefont {J.~T.}\ \bibnamefont {Omotani}}, \bibinfo {author} {\bibfnamefont {M.~V.}\ \bibnamefont {Umansky}}, \bibinfo {author} {\bibfnamefont {D.}~\bibnamefont {Del~Sorbo}}, \bibinfo {author} {\bibfnamefont {B.}~\bibnamefont {Dudson}}, \bibinfo {author} {\bibfnamefont {J.~T.}\ \bibnamefont {Parker}}, \bibinfo {author} {\bibfnamefont {G.~D.}\ \bibnamefont {Kerbel}}, \bibinfo {author} {\bibfnamefont {M.}~\bibnamefont {Sherlock}}, \ and\ \bibinfo {author} {\bibfnamefont {C.~P.}\ \bibnamefont {Ridgers}},\ }\bibfield  {title} {\enquote {\bibinfo {title} {{Testing nonlocal models of electron thermal conduction for magnetic and inertial
  confinement fusion applications}},}\ }\href {\doibase 10.1063/1.5001079} {\bibfield  {journal} {\bibinfo  {journal} {Physics of Plasmas}\ }\textbf {\bibinfo {volume} {24}},\ \bibinfo {pages} {092309} (\bibinfo {year} {2017})},\ \Eprint {http://arxiv.org/abs/https://pubs.aip.org/aip/pop/article-pdf/doi/10.1063/1.5001079/15654563/092309\_1\_online.pdf} {https://pubs.aip.org/aip/pop/article-pdf/doi/10.1063/1.5001079/15654563/092309\_1\_online.pdf} \BibitemShut {NoStop}%
\bibitem [{\citenamefont {Ma}\ \emph {et~al.}(2022)\citenamefont {Ma}, \citenamefont {Patel}, \citenamefont {Sherlock}, \citenamefont {Farmer},\ and\ \citenamefont {Johnsen}}]{Ma2022}%
  \BibitemOpen
  \bibfield  {author} {\bibinfo {author} {\bibfnamefont {K.~H.}\ \bibnamefont {Ma}}, \bibinfo {author} {\bibfnamefont {M.~V.}\ \bibnamefont {Patel}}, \bibinfo {author} {\bibfnamefont {M.}~\bibnamefont {Sherlock}}, \bibinfo {author} {\bibfnamefont {W.~A.}\ \bibnamefont {Farmer}}, \ and\ \bibinfo {author} {\bibfnamefont {E.}~\bibnamefont {Johnsen}},\ }\bibfield  {title} {\enquote {\bibinfo {title} {Thermal transport modeling of laser-irradiated spheres},}\ }\href {\doibase 10.1063/5.0005552} {\bibfield  {journal} {\bibinfo  {journal} {Physics of Plasmas}\ }\textbf {\bibinfo {volume} {29}},\ \bibinfo {pages} {112307} (\bibinfo {year} {2022})},\ \Eprint {http://arxiv.org/abs/https://pubs.aip.org/aip/pop/article-pdf/doi/10.1063/5.0005552/16627725/112307\_1\_online.pdf} {https://pubs.aip.org/aip/pop/article-pdf/doi/10.1063/5.0005552/16627725/112307\_1\_online.pdf} \BibitemShut {NoStop}%
\bibitem [{\citenamefont {Marocchino}\ \emph {et~al.}(2013)\citenamefont {Marocchino}, \citenamefont {Tzoufras}, \citenamefont {Atzeni}, \citenamefont {Schiavi}, \citenamefont {Nicolaï}, \citenamefont {Mallet}, \citenamefont {Tikhonchuk},\ and\ \citenamefont {Feugeas}}]{Marocchino}%
  \BibitemOpen
  \bibfield  {author} {\bibinfo {author} {\bibfnamefont {A.}~\bibnamefont {Marocchino}}, \bibinfo {author} {\bibfnamefont {M.}~\bibnamefont {Tzoufras}}, \bibinfo {author} {\bibfnamefont {S.}~\bibnamefont {Atzeni}}, \bibinfo {author} {\bibfnamefont {A.}~\bibnamefont {Schiavi}}, \bibinfo {author} {\bibfnamefont {P.~D.}\ \bibnamefont {Nicolaï}}, \bibinfo {author} {\bibfnamefont {J.}~\bibnamefont {Mallet}}, \bibinfo {author} {\bibfnamefont {V.}~\bibnamefont {Tikhonchuk}}, \ and\ \bibinfo {author} {\bibfnamefont {J.-L.}\ \bibnamefont {Feugeas}},\ }\bibfield  {title} {\enquote {\bibinfo {title} {{Comparison for non-local hydrodynamic thermal conduction models}},}\ }\href {\doibase 10.1063/1.4789878} {\bibfield  {journal} {\bibinfo  {journal} {Physics of Plasmas}\ }\textbf {\bibinfo {volume} {20}},\ \bibinfo {pages} {022702} (\bibinfo {year} {2013})},\ \Eprint {http://arxiv.org/abs/https://pubs.aip.org/aip/pop/article-pdf/doi/10.1063/1.4789878/16146337/022702\_1\_online.pdf}
  {https://pubs.aip.org/aip/pop/article-pdf/doi/10.1063/1.4789878/16146337/022702\_1\_online.pdf} \BibitemShut {NoStop}%
\bibitem [{\citenamefont {Lindl}(1995{\natexlab{b}})}]{lindlICF}%
  \BibitemOpen
  \bibfield  {author} {\bibinfo {author} {\bibfnamefont {J.}~\bibnamefont {Lindl}},\ }\bibfield  {title} {\enquote {\bibinfo {title} {Development of the indirect‐drive approach to inertial confinement fusion and the target physics basis for ignition and gain},}\ }\href {\doibase 10.1063/1.871025} {\bibfield  {journal} {\bibinfo  {journal} {Physics of Plasmas}\ }\textbf {\bibinfo {volume} {2}},\ \bibinfo {pages} {3952,3996} (\bibinfo {year} {1995}{\natexlab{b}})},\ \Eprint {http://arxiv.org/abs/https://pubs.aip.org/aip/pop/article-pdf/2/11/3933/19277171/3933\_1\_online.pdf} {https://pubs.aip.org/aip/pop/article-pdf/2/11/3933/19277171/3933\_1\_online.pdf} \BibitemShut {NoStop}%
\bibitem [{\citenamefont {Sherlock}\ and\ \citenamefont {Bissell}(2020)}]{SherlockBissel2020}%
  \BibitemOpen
  \bibfield  {author} {\bibinfo {author} {\bibfnamefont {M.}~\bibnamefont {Sherlock}}\ and\ \bibinfo {author} {\bibfnamefont {J.~J.}\ \bibnamefont {Bissell}},\ }\bibfield  {title} {\enquote {\bibinfo {title} {Suppression of the biermann battery and stabilization of the thermomagnetic instability in laser fusion conditions},}\ }\href {\doibase 10.1103/PhysRevLett.124.055001} {\bibfield  {journal} {\bibinfo  {journal} {Phys. Rev. Lett.}\ }\textbf {\bibinfo {volume} {124}},\ \bibinfo {pages} {055001} (\bibinfo {year} {2020})}\BibitemShut {NoStop}%
\bibitem [{\citenamefont {Bell}(1983{\natexlab{b}})}]{BellESTest}%
  \BibitemOpen
  \bibfield  {author} {\bibinfo {author} {\bibfnamefont {A.~R.}\ \bibnamefont {Bell}},\ }\bibfield  {title} {\enquote {\bibinfo {title} {{Electron energy transport in ion waves and its relevance to laser‐produced plasmas}},}\ }\href {\doibase 10.1063/1.864018} {\bibfield  {journal} {\bibinfo  {journal} {The Physics of Fluids}\ }\textbf {\bibinfo {volume} {26}},\ \bibinfo {pages} {279--284} (\bibinfo {year} {1983}{\natexlab{b}})},\ \Eprint {http://arxiv.org/abs/https://pubs.aip.org/aip/pfl/article-pdf/26/1/279/12262372/279\_1\_online.pdf} {https://pubs.aip.org/aip/pfl/article-pdf/26/1/279/12262372/279\_1\_online.pdf} \BibitemShut {NoStop}%
\bibitem [{\citenamefont {Kingham}\ and\ \citenamefont {Bell}(2004)}]{KinghamVFP}%
  \BibitemOpen
  \bibfield  {author} {\bibinfo {author} {\bibfnamefont {R.}~\bibnamefont {Kingham}}\ and\ \bibinfo {author} {\bibfnamefont {A.}~\bibnamefont {Bell}},\ }\bibfield  {title} {\enquote {\bibinfo {title} {{An implicit Vlasov–Fokker–Planck code to model non-local electron transport in 2-D with magnetic fields}},}\ }\href {\doibase https://doi.org/10.1016/j.jcp.2003.08.017} {\bibfield  {journal} {\bibinfo  {journal} {Journal of Computational Physics}\ }\textbf {\bibinfo {volume} {194}},\ \bibinfo {pages} {1--34} (\bibinfo {year} {2004})}\BibitemShut {NoStop}%
\bibitem [{\citenamefont {Bobylev}\ and\ \citenamefont {Chuyanov}(1976)}]{Bobylev}%
  \BibitemOpen
  \bibfield  {author} {\bibinfo {author} {\bibfnamefont {A.}~\bibnamefont {Bobylev}}\ and\ \bibinfo {author} {\bibfnamefont {V.}~\bibnamefont {Chuyanov}},\ }\bibfield  {title} {\enquote {\bibinfo {title} {On the numerical solution of {L}andau's kinetic equation},}\ }\href {\doibase https://doi.org/10.1016/0041-5553(76)90109-9} {\bibfield  {journal} {\bibinfo  {journal} {USSR Computational Mathematics and Mathematical Physics}\ }\textbf {\bibinfo {volume} {16}},\ \bibinfo {pages} {121--130} (\bibinfo {year} {1976})}\BibitemShut {NoStop}%
\end{thebibliography}%

\end{document}